\documentclass[a4paper,11pt]{article}
\pdfoutput=1 

\usepackage{jinstpub} 
\usepackage{multirow}
\usepackage{xspace}
\newcommand{\legend} {{\textsc{Legend}}\xspace}
\usepackage{lineno}

\title{Optical properties of low background PEN structural components for the \legend-$200$ experiment}

\makeatletter
\newcommand\footnoteref[1]{\protected@xdef\@thefnmark{\ref{#1}}\@footnotemark , }
\makeatother

\author[a,1]{L.~Manzanillas,\note{Corresponding author.}} 
\author[c]{Y.~Efremenko,}
\author[b]{M.~Febbraro,}
\author[a]{F.~Fischer,}
\author[a]{M.~Guitart Corominas,}
\author[c]{B.~Hackett,}
\author[d]{A.~Leonhardt,}
\author[a]{B.~Majorovits,}
\author[a]{O.~Schulz,}

\affiliation[a]{Max-Planck-Institut  f\"ur  Physik,  80805  Munich,  Germany}
\affiliation[b]{Oak Ridge National Laboratory, Oak Ridge, Tennessee 37830, USA}
\affiliation[c]{Department of Physics and Astronomy, University of Tennessee, Knoxville, Tennessee 37916, USA}
\affiliation[d]{Physik Department, Technische Universität, München, Germany}

\emailAdd{manzanil@mpp.mpg.de}

\abstract{Polyethylene Naphthalate (PEN) plastic scintillator has been identified as potential self-vetoing structural material in low-background physics experiments. Radio-pure scintillating components have been produced from PEN using injection compression molding technology. These low-background PEN components will be used as optically active holders to mount the Germanium detectors in the \legend-$200$ neutrinoless double beta decay experiment. In this paper, we present the measurement of the optical properties of these PEN components. The scintillation light emission spectrum, time constant, attenuation and bulk absorption length as well as light output and light yield are reported. In addition, the surface of these PEN components has been characterized and an estimation of the surface roughness is presented. The light output of the final \legend-$200$ detector holders has been measured and is reported. These measurements were used to estimate the self-vetoing efficiency of these holders.}


\begin{document}
\maketitle
\flushbottom

\section{Introduction}
\label{sec:intro}

Rare event physics experiments such as dark matter or neutrinoless double beta  ($0\nu\beta\beta$) decay searches demand ultra-low backgrounds. Hence, ultra-pure materials are required for the structural materials and for the detectors themselves. 
Additional strategies to mitigate external backgrounds include active vetoes around the detectors.
In the GERmanium Detector Array ({\textsc{Gerda}}) experiment, the predecessor of the Large Enriched Germanium Experiment for Neutrinoless $\beta\beta$ Decay (\legend), liquid Argon (LAr) was used to cool down the Germanium (Ge) detectors and at the same time it also served as an active shielding. Interactions of charged particles in LAr produce Vacuum Ultra-Violet (VUV) light ($\sim127$ nm) \cite{Heindl:2010zz} that is used to veto external backgrounds.
However, the 127 nm VUV light from LAr cannot be efficiently detected using silicon photomultipliers (SiPMs),  multi-pixel photon counters (MPPCs), or photomultiplier tubes (PMTs).  The common solution to this problem is the use of wavelength-shifting (WLS) coatings,  which absorb the VUV photons and then re-emit light of wavelengths at which standard photo-sensors are more efficient. 
Thus, in {\textsc{Gerda}}, the  VUV light from LAr was collected using a curtain of WLS and scintillating optical fibers placed around the detector array \cite{Agostini:2017hit}. These optical fibers guide the photons that are then collected by SiPMs placed on top of the setup. This method has proved to be very efficient in discriminating external backgrounds. Nevertheless, in order to increase the sensitivity of the next generation of  $0\nu\beta\beta$-decay Ge experiments, the backgrounds produced around the Ge detectors need to be suppressed even more. This new generation of experiments, \legend-$200$ and \legend-$1000$, are currently being developed  by the \legend collaboration. Increasing the light collection efficiency of events produced in the vicinity of the Ge detectors is of paramount importance to reach the \legend physics goals. The \legend-$1000$ experiment is designed to probe the $0\nu\beta\beta$-decay  with a 99.7\%-CL discovery sensitivity in the $^{76}$Ge half-life of $1.3\times10^{28}$ years, corresponding to an effective Majorana mass upper limit in the range of 9-21 meV, to cover the inverted-ordering neutrino mass scale with 10 years of live time \cite{LEGEND:2021bnm}. 

The support structures used to mount the Ge detectors usually consist of optically inactive and nontransparent materials.
The design of these support structures has been optimized to reduce the amount of inactive materials. In addition to being a potential background source, the structural materials also absorb the scintillation light of LAr, decreasing the background identification capabilities for events originating in the vicinity of the detectors. As part of the \legend R\&D program, radio-pure polyethylene-naphthalate (PEN) structures have been produced using injection compression molding technology \cite{Efremenko:2019xbs,Efremenko:2021olf}.
PEN is a commercially-available polyester, which has a yield strength higher than copper at cryogenic temperatures and therefore can be used as structural material in cryogenic physics experiments. Moreover, it can act as a wavelength shifter and it scintillates in the blue region between 410 nm and 550 nm, which is ideal for most of photo-sensors devices \cite{Kuzniak:2020oka,Abraham:2021otn,Kuzniak:2018dcf,Araujo:2021buv}.

A structural material with both scintillation and WLS capabilities such as PEN, in an environment like the \legend-$200$/\legend-$1000$ setups has several advantages. 
Thanks to the scintillation, backgrounds originated by radio-impurities within the bulk PEN material can produce a light signal that could be used to identify these events. This self-vetoing capability combined with a high radiopurity can strongly suppress the backgrounds induced by PEN.
In the same way external backgrounds depositing some energy in PEN can also be identified.
In addition, PEN will improve the LAr light collection efficiency thanks to the WLS capabilities.  
Consequently, a higher identification of backgrounds generated in the vicinity of the detectors can be achieved. 
The PEN efficiency to shift the wavelength of LAr light ($\sim$127 nm) to blue light ($\sim$450 nm) is under investigation. Results vary from 12\% to 70\% compared to tetraphenyl butadiene (TPB) and are dependent on sample and setup \cite{Araujo:2021buv,Abraham:2021otn,Boulay:2021njr}.

Since PEN is a scintillating material being used as an optically active structural material for the first time in the \legend-$200$ experiment, most of its optical properties are as yet unknown. A precise knowledge of the PEN optical properties of the Ge detector holders used in the \legend-$200$ experiment is of paramount importance in order to determine  the background rejection efficiency for different radiation sources.
Hence, the optical properties of these PEN structures such as emission spectrum, time constant, attenuation, bulk absorption length as a function of wavelength and light yield have to be carefully studied and characterized. 
The knowledge of these parameters is crucial to allow the determination of the detection efficiencies of backgrounds originating in the close surroundings of the Ge detectors, for example $^{42}$Ar in the liquid argon or in the support structures in the \legend experiment.
Knowledge of these parameters will allow to estimate the expected background contributions in the future \legend-$1000$ experiment. Furthermore, these results can be used to optimize the geometry of the mounting structures to maximize light collection.


This paper is organized as follows. The PEN holders designed for \textsc{Legend}-200 and the setups for their optical characterization are described in sections \ref{S:holders} and \ref{S:setup}.  The surface and optical characterization of the PEN production for \textsc{Legend}-200 are discussed in section \ref{sec:surface} and \ref{S:scintillation}, respectively. Finally, the light output of the PEN holders as well as estimations on the light output and photon detection efficiency are presented in section \ref{S:efficiency}. 
\section{Low-background PEN scintillating structures in the \legend-$200$ experiment}
\label{S:holders}

\begin{figure}[h]
    \centering
    \includegraphics[width=0.4\textwidth]{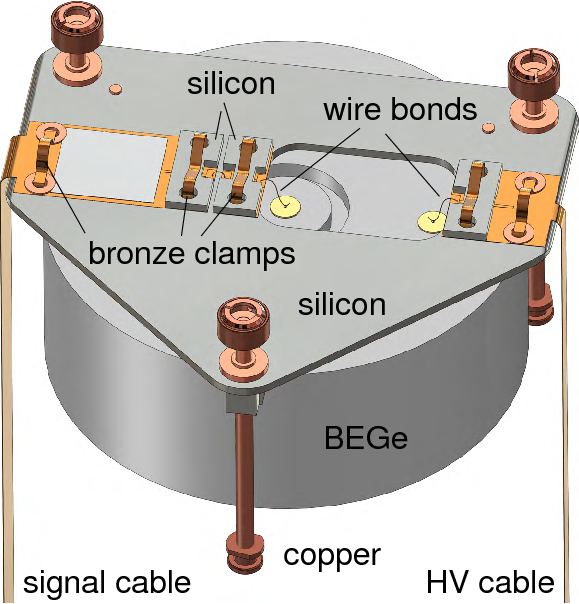}
    \scalebox{-1}[1]{\includegraphics[trim=120 250 50 250, clip, width=0.45\textwidth ]{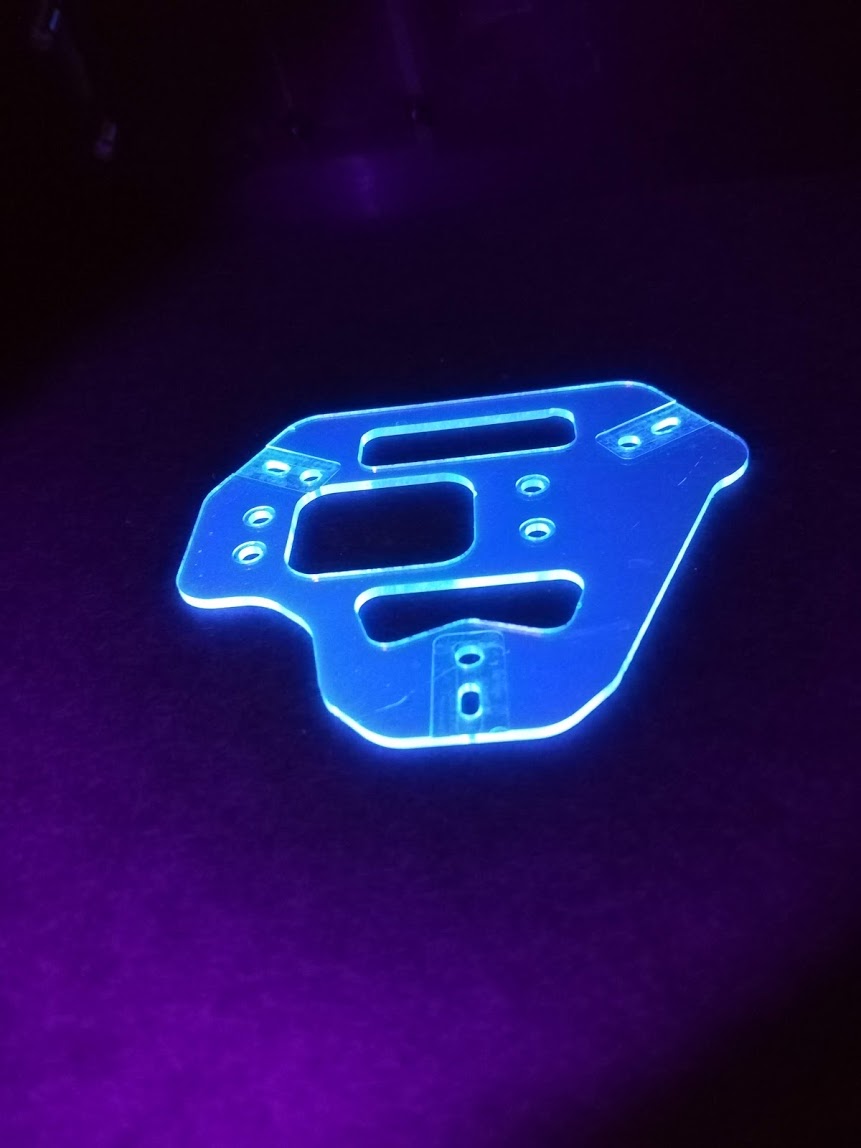}}
    \caption{Left: Sketch of the support structures used to mount the Ge detectors in {\textsc{Gerda}}. Right:  Scintillating PEN holder for \legend-$200$ excited with a UV lamp. The PEN holders will replace the high purity Si plates used previously in the {\textsc{Gerda}} experiment (left panel). }
    \label{fig:jig}
\end{figure}

Scintillating low-background structures were produced by employing injection compression molding technology using commercially available  PEN granulate (TN-8065 SC) from Teijin-DuPont. A detailed description of the whole process from granulate to final radio-pure  and optically active support holders can be found in reference \cite{Efremenko:2021olf}.
The typical support structures for the germanium detectors used in the {\textsc{Gerda}} experiment are shown in Figure \ref{fig:jig}. These structures included copper elements and a high purity silicon (Si) plate that was used to carry the required electronics to supply high voltage (HV) and to readout the signals from the detectors as well as other components needed to mount the Ge detectors. 
Contamination in the close detector surrounding or on the detector surface were the most prominent backgrounds in {\textsc{Gerda}} \cite{Abramov:2019hhh}. Hence, new strategies to mitigate these backgrounds must be developed.

In the \legend-$200$ experiment, optically active detector support holders will replace the high-purity silicon plates previously used in the {\textsc{Gerda}} experiment (see Figure \ref{fig:jig}). 
The active PEN holders will complement the LAr veto system allowing to minimize non-active components around the Ge detectors.
As a consequence of the scintillating and WLS capabilities of PEN, an improved light collection and identification of external backgrounds produced in between the Ge detectors can be obtained. 

Finally, in order to minimize the mass of the PEN holders, mechanical simulations were performed to optimize their design. These studies produced three optimized designs to mount all types of Ge detectors used in the \legend-$200$ experiment. 
The final design of the low-mass PEN holders  was validated during the \legend-$200$ prototyping tests at LNGS in 2020 \cite{Abt:2020pwk,Efremenko:2021olf,Manzanillas:2022pat}.

\section{Test bench setups for optical characterization}
\label{S:setup}
Aiming to assess the most important optical parameters like the light yield, light output and attenuation length of the molded radio-pure \legend-$200$ PEN scintillator, several test benches were set up. These setups included two dark boxes with a spectrometer and several photo-multiplier tubes (PMTs) with which the PEN optical properties were measured.
In order to guarantee reproducible measurements, 3D printed support structures featuring 0.1~mm precision were used, ensuring that the scintillator samples together with the photo-sensor devices could be arranged with a precision of 0.1~mm.   

\begin{figure}[h]
    \centering
    \includegraphics[width=0.37\textwidth ]{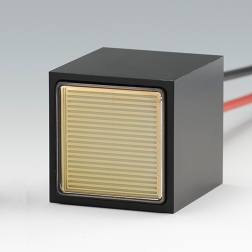}
    \includegraphics[width=0.55\textwidth]{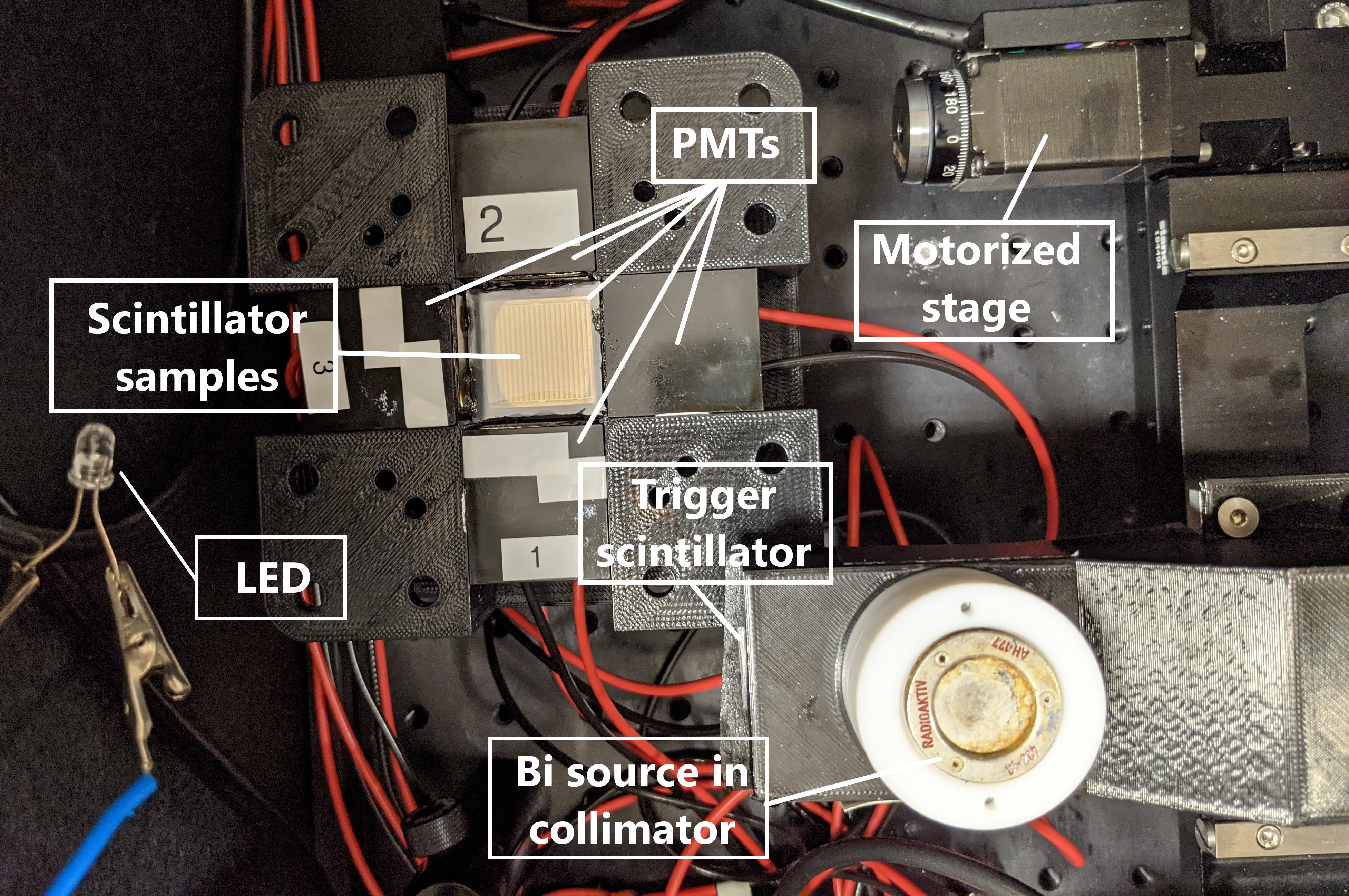}
    \caption{Left: photo-multiplier tube assembly of type H11934-300 used for this study. Right: Interior of the dark box used to measure the optical properties of PEN. A radioactive $^{207}$Bi source with a collimator (bottom right) is mounted on a motorized stage. A 3D-printed holder is used to mount the PEN sample with the PMTs. }
    \label{fig:setup}
\end{figure}

The PMTs used for these measurements are of type H11934-300 from  HAMAMATSU photonics (see Figure \ref{fig:setup}). Each PMT assembly provides a sensitive square area of around 23$\times$23 mm$^{2}$. A maximum quantum efficiency of $\eta=39\%$ at 420 nm is quoted by the manufacturer \cite{hamamatsuPMT}. 
The PMT window material is made of borosilicate glass, while the photocathode material is of type  EGBA (Extended green bialkali).

A triggering mechanism was developed to select beta particles from a $^{207}$Bi radioactive source. This mechanism is described in subsection \ref{section_trigger}.
Figure \ref{fig:setup} (Right) shows a photo of the setup employed for this research. It features the $^{207}$Bi source placed on a collimator vertically above a PEN sample coupled to PMTs in a 3D printed holder. The source together with the trigger system were mounted on a remote-controlled stage that can be moved two-dimensionally in a parallel layer 15 mm above the PEN sample. This 8MTF-102LS05 stage from STANDA features micrometer resolution and guarantees equidistant steps from point to point with an uncertainty of $2.5\:\mathrm{\mu m}$.

To study the light emission properties of PEN, an Andor Shamrock 193i spectrograph was employed. PEN samples were optically coupled to it and exposed to photons from a 345~nm light-emitting diode~(LED). The bulk absorption length of PEN was assessed with direct and indirect methods encompassing measurements and Geant4 optical simulations. To this end, a Shimadzu UV-2700i UV-Vis Spectrophotometer and a Lambda 850 UV-VIS spectrometer from Perkin Elmer were used.

\subsection{The Bi-207 radioactive source}
\label{subsection-Bi-source}
\begin{figure}[h]
    \centering
    \includegraphics[width=0.7\textwidth]{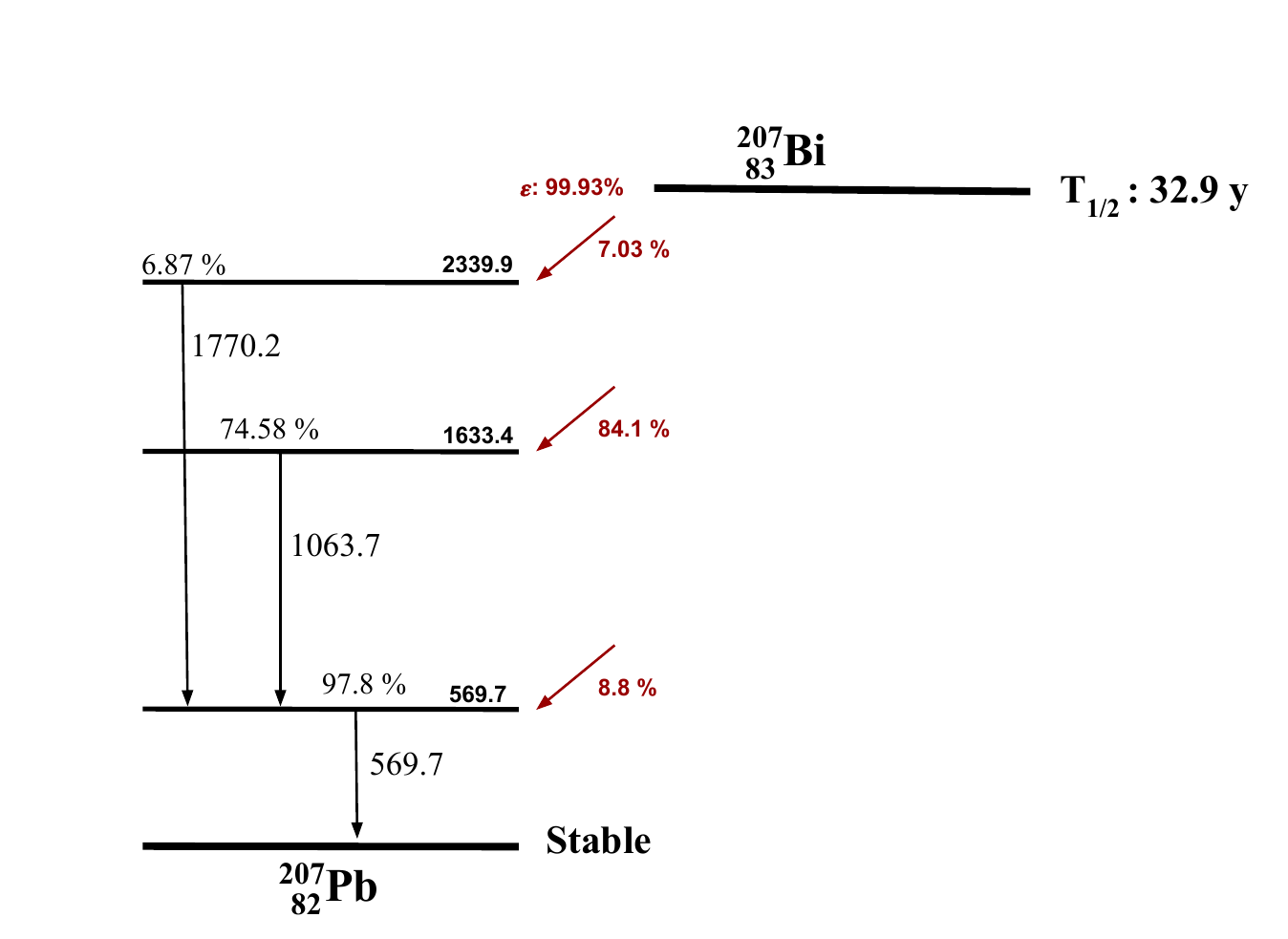}
    \caption{The three major $\gamma$ lines of the $^{207}$Bi radioactive source.}
    \label{fig:Bi207}
\end{figure}

Under the operational conditions of the \legend-$200$ experiment, most of the particle interactions in the PEN holders will have their origin from natural radioactivity within the holders or in their vicinity. The most relevant of these particles have energies at the MeV scale. Hence, the optical properties must be characterized using radiation of similar energies. Thus, a $^{207}$Bi radioactive source is well suited for this purpose. The $^{207}$Bi decays through electron capture (99.9\%) into excited states of $^{207}$Pb. The excited $^{207}$Pb goes into its ground state through gamma emissions or through emission of conversion electrons. There are three major gamma lines with energies of 570, 1064 and 1770~keV \cite{PhysRev.99.695} as illustrated in Figure~\ref{fig:Bi207}. These $\gamma$-ray emissions can be replaced by atomic K, L or M shell conversion electrons as presented in Table~\ref{table:1}. Most of the useful conversion electrons are attributed to the 1064~keV de-excitation and have an energy between 976 and 1060~keV with a total probability of 9.5~\%. In contrast to the Compton continuum produced by gammas, these mono-energetic electrons can be absorbed in a few millimeters of PEN or materials with similar density, providing a well defined energy deposit.
\begin{table}[h]
\centering
\begin{tabular}{|c | c | c | c|} 
 \hline
 Transition [keV] & Shell & Energy [keV] & Probability [\%]\\  
 \hline
 \multirow{ 3}{*}{569.7} & K & 482 & 1.55 \\ 
  & L & 555 & 0.43 \\
  & M & 566 & 0.11 \\
 \hline
 \multirow{ 4}{*}{1063.7} & K & 976 & 7.11 \\
  & L & 1049 & 1.84 \\ 
  & M & 1060 & 0.44 \\
  & N & 1063 & 0.12 \\
 \hline
 1770.2 & K & 1682 & 0.02 \\
 \hline
\end{tabular}
\caption{Main conversion electrons from $^{207}$Bi \cite{PhysRev.99.695,tableisotopes}.}
\label{table:1}
\end{table}

The $^{207}$Bi source from Eckert~\&~Ziegler had an activity of about 170~kBq.
The source was fabricated by gluing a thin deposit of active material between two titanium foils, each 2.4~mg/cm$^2$ resulting in a thickness of 5.3~$\mu$m. The foils are supported in an aluminium frame of 25~mm in diameter, 3~mm thick. The diameter of the active deposit is approximately 5~mm.

\subsection{Trigger system for electron selection}\label{section_trigger}
\begin{figure}[h]
    \centering
    \includegraphics[width=0.4\textwidth]{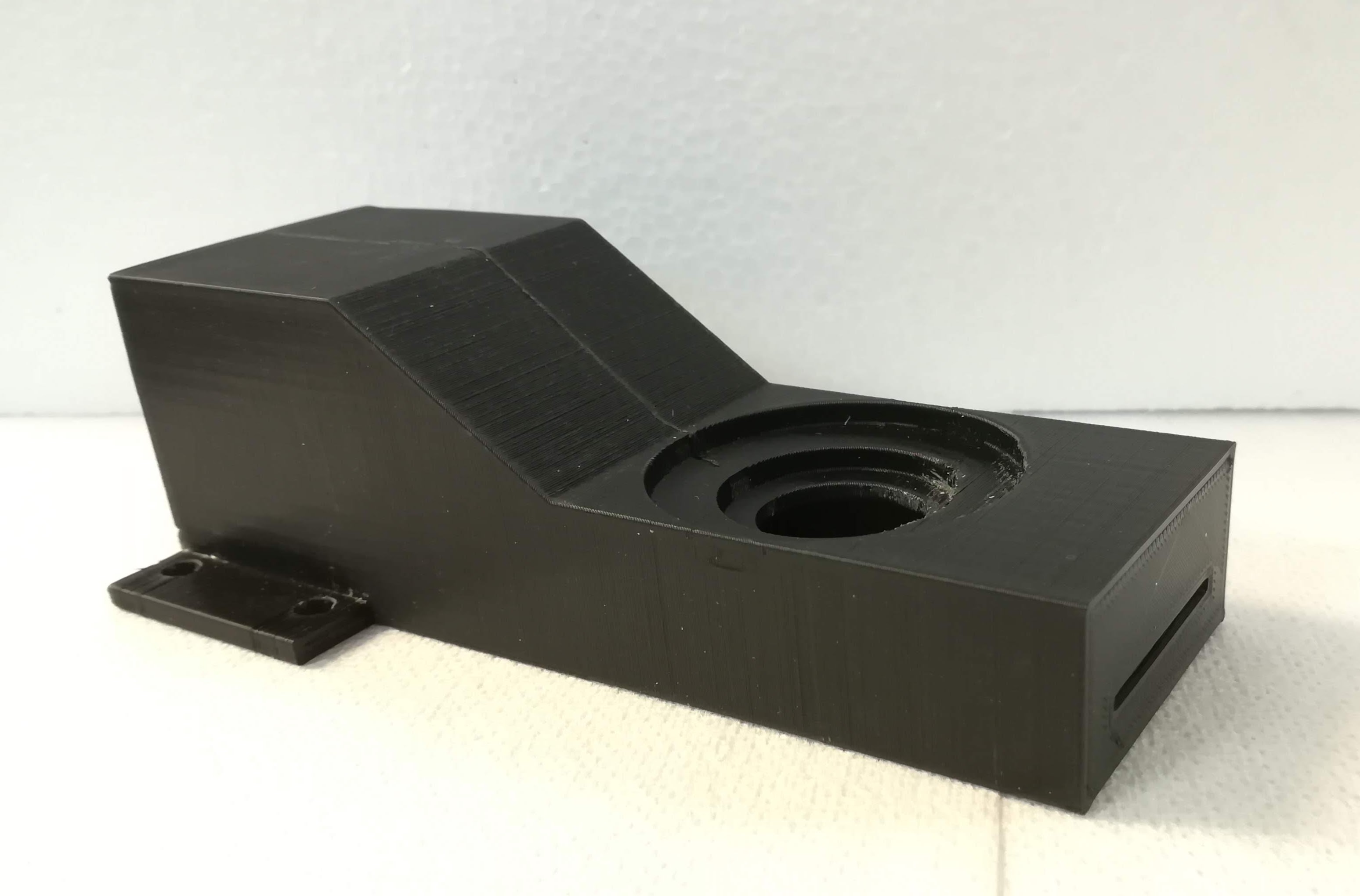}
    \includegraphics[width=0.55\textwidth ]{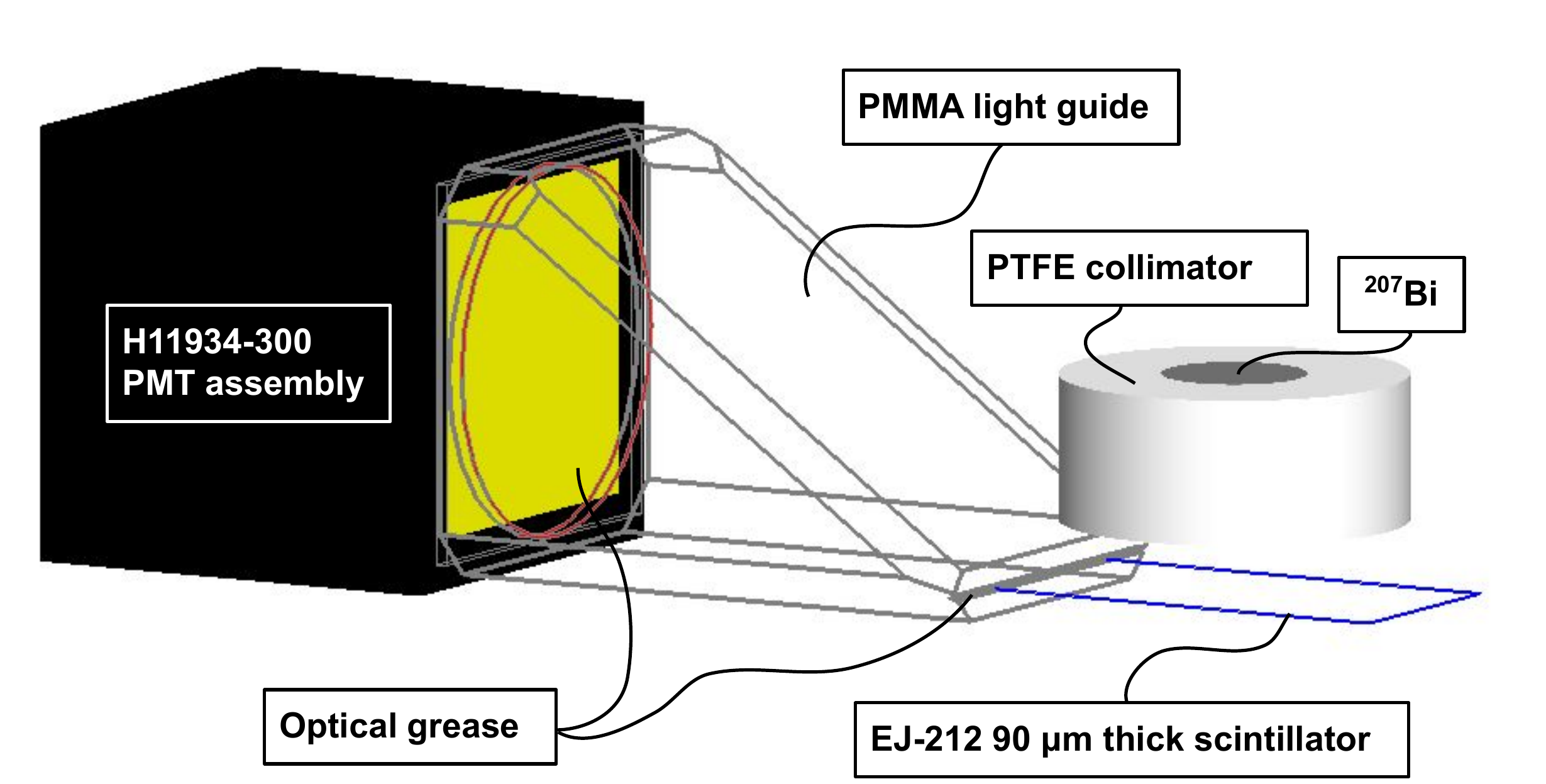}
    \caption{Left: 3D printed support structure used to mount the trigger setup. Right: Device used to select electrons from the $^{207}$Bi source using a 90 $\mu$m thick scintillator.}
    \label{fig:triggersystem}
\end{figure}

In order to select only the "mono-"energetic electrons from the $^{207}$Bi source, a trigger system inspired from reference \cite{Abreu:2018ajc} was developed. 
The trigger system consisted of a thin EJ-212 plastic scintillator (90~$\mu$m thick) from ELJEN TECHNOLOGY, that was coupled to a  polymethyl methacrylat (PMMA) light guide, which in turn was coupled to a 1~inch PMT as is sketched in Figure~\ref{fig:triggersystem}~(Right). Optical EJ-550 grease was used to guarantee a good optical coupling between the EJ-212 scintillator and the PMMA light guide and between the latter and the PMT. This ensemble was then placed in a 3D-printed black support as shown in Figure~\ref{fig:triggersystem}~(Left). 

The polymer base of the EJ-212 scintillator is Polyvinyltoluene (PVT) and has a density of 1.023~g/cm$^{3}$. The reported light yield by the manufacturer is $10000 \pm 200$ photons per MeV, with a spectrum peaking at 423~nm. This EJ-212 scintillator has a decay time of 2.5~ns, while the light attenuation length is reported to be 250~cm \cite{eljen}.
Electrons going through the EJ-212 plastic scintillator will deposit energy and excite the material, initiating a scintillation light signal. The mean deposited energy by electrons traversing the EJ-212 scintillator was estimated using Geant4 simulations. It was found that electrons from the $^{207}$Bi source traversing the EJ-212 scintillator on average deposit around 25~keV. This amount of energy will generate about 250 optical photons of which some can be detected by the PMT of this system. 
The electron selection was performed by asking for the detection of at least 1 photon in the PMT of the trigger system in coincidence with a signal in one of the PMTs surrounding the PEN samples.

Finally, electrons were collimated by placing a PTFE collimator between the $^{207}$Bi source and the EJ-212 scintillator. This collimator consisted of a cylinder with an internal and external diameter of 2 and 20~mm, respectively, and a height of 20~mm. 
In this configuration only the collimated electrons passing through the collimator hole  interact with the EJ-200 scintillator. 

\subsection{Data taking and processing}

The signals from the PMTs were readout using a Struck SIS3316 digitizer. A threshold trigger of around 0.5 photo-electrons was used in the PMT of the trigger system described in the previous section. For each triggered event, signals sampled at 250~MHz and containing complete waveforms with 128 samples (512~ns) in each PMT were readout for post-processing and analysis.  

Signals were processed offline using the Julia programming language \cite{bezanson2017julia}. After baseline subtraction the charge was converted to the number of detected photons using the single photo-electron gain response of each PMT. The single photo-electron conversion factor was calibrated for each PMT and is described in detail in the next section. Finally, the total number of detected photons was computed by adding the number of detected photons of all PMTs coupled to the scintillator sample.

\subsection{PMT single photo-electron gain calibration}
\label{subsec:pmt-calibration}
The PMTs used for this study were calibrated using light emitting diodes (LED) of 400 and 450~nm. The LEDs were operated at low intensity using a pulse generator with a frequency of 1~kHz. 
The trigger threshold was set below 0.5 photo-electrons in order to capture the complete single photo-electron (SPE) distribution.

The SPE selection consisted of finding the number of peaks in the waveform above a given threshold after baseline subtraction.
Then, only waveforms with a single peak were used to compute the SPE charge by integrating the charge in the last 64 samples, while the pedestal was computed using the first 64 samples of the waveform. A pre-trigger window of 90 samples was used in order to guarantee the position of the SPE peak in the last 64 samples.        
The charge distribution thus obtained was then fitted using the following model \cite{Bauer:2011ne}
\begin{equation}\label{eq:spe}
\setlength{\jot}{10pt}
\begin{aligned}
f (x) &= \underbrace{\dfrac{N_{\text{Ped}}}{\sigma_{\text{Ped}}\sqrt{2\pi}}\text{ exp}\left({-\dfrac{(x-\mu_{\text{Ped}})^2}{2\sigma_{\text{Ped}}^2}}\right)}_{\text{pedestal}} +   \underbrace{\dfrac{N_{\text{Exp}}}{\tau}\text{ exp}\left({-\dfrac{x}{\tau}}\right)}_{\text{ badly amplified events}}\\
&+\underbrace{\dfrac{N_{\text{1PE}}}{\sigma_{\text{1PE}}\sqrt{2\pi}}\text{ exp}\left({-\dfrac{(x-\mu_{\text{1PE}})^2}{2\sigma_{\text{1PE}}^2}}\right)}_{\text{first PE}}
+\underbrace{\dfrac{N_{\text{2PE}}}{\sigma_{\text{2PE}}\sqrt{2\pi}}\text{ exp}\left({-\dfrac{(x-2\mu_{\text{1PE}})^2}{2\sigma_{\text{2PE}}^2}}\right)}_{\text{second PE}}
\end{aligned}
\end{equation}
where $N$, $\mu$ and $\sigma$ correspond to the number of events, the position and width of the individual Gaussian distributions, respectively. Here, the first Gaussian term corresponds to the pedestal, while the second and third correspond to the first and second photo-electron peaks, respectively. In order to take the badly amplified events\footnote{In these events an electron released from the photocathode may follow a non-ideal trajectory which will result in secondary electrons potentially not reaching the next stage of amplification. This will ultimately result in lower amplification and is caused by electric field imperfections in the PMT \cite{Bauer:2011ne,Anthony_2018}.} into account, an exponential term with decay constant $\tau$ was used, which is an empirical function to describe these events.

\begin{figure}[h]
    \centering
    \includegraphics[width=0.45\textwidth]{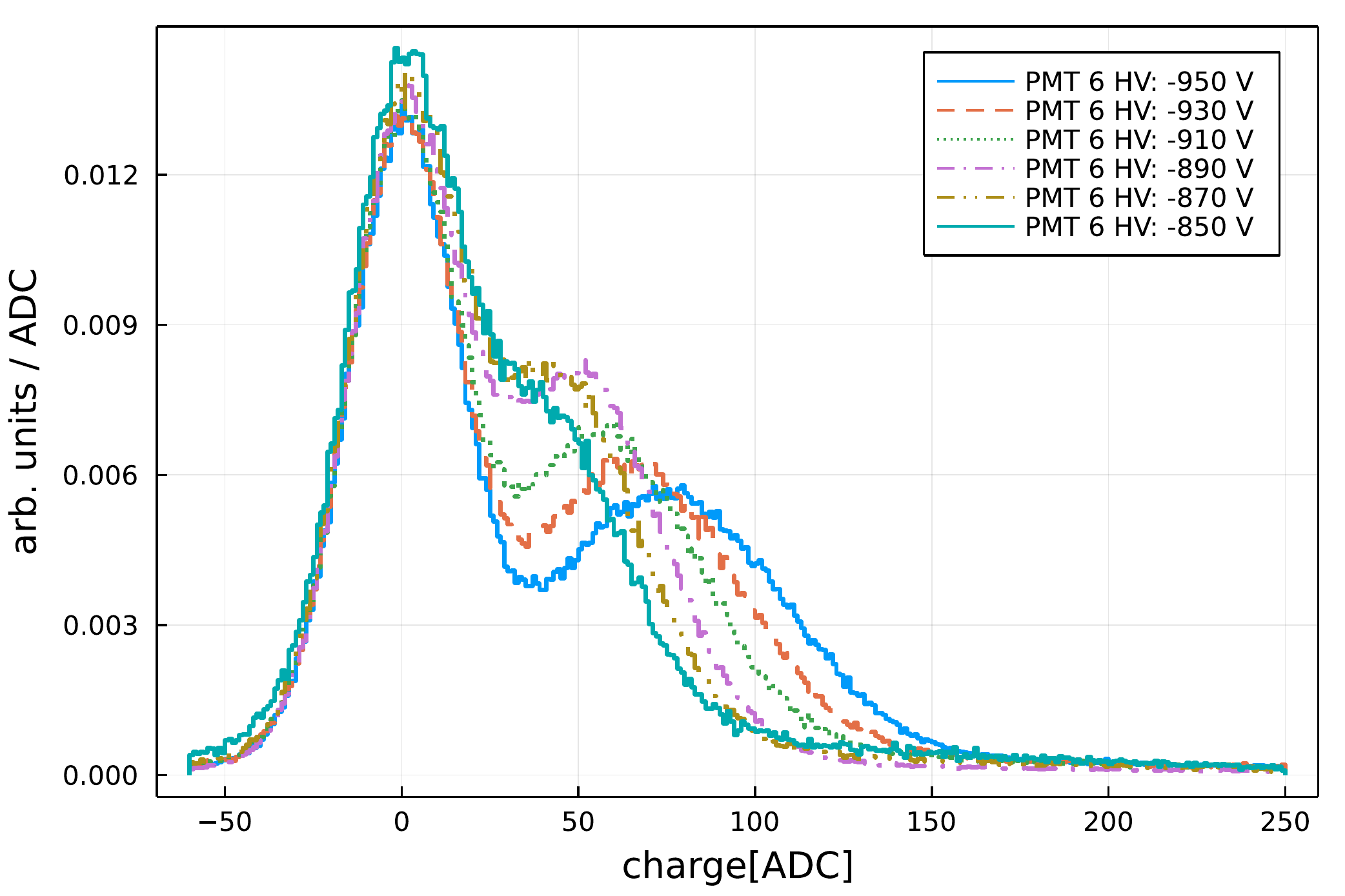}
    \includegraphics[width=0.45\textwidth ]{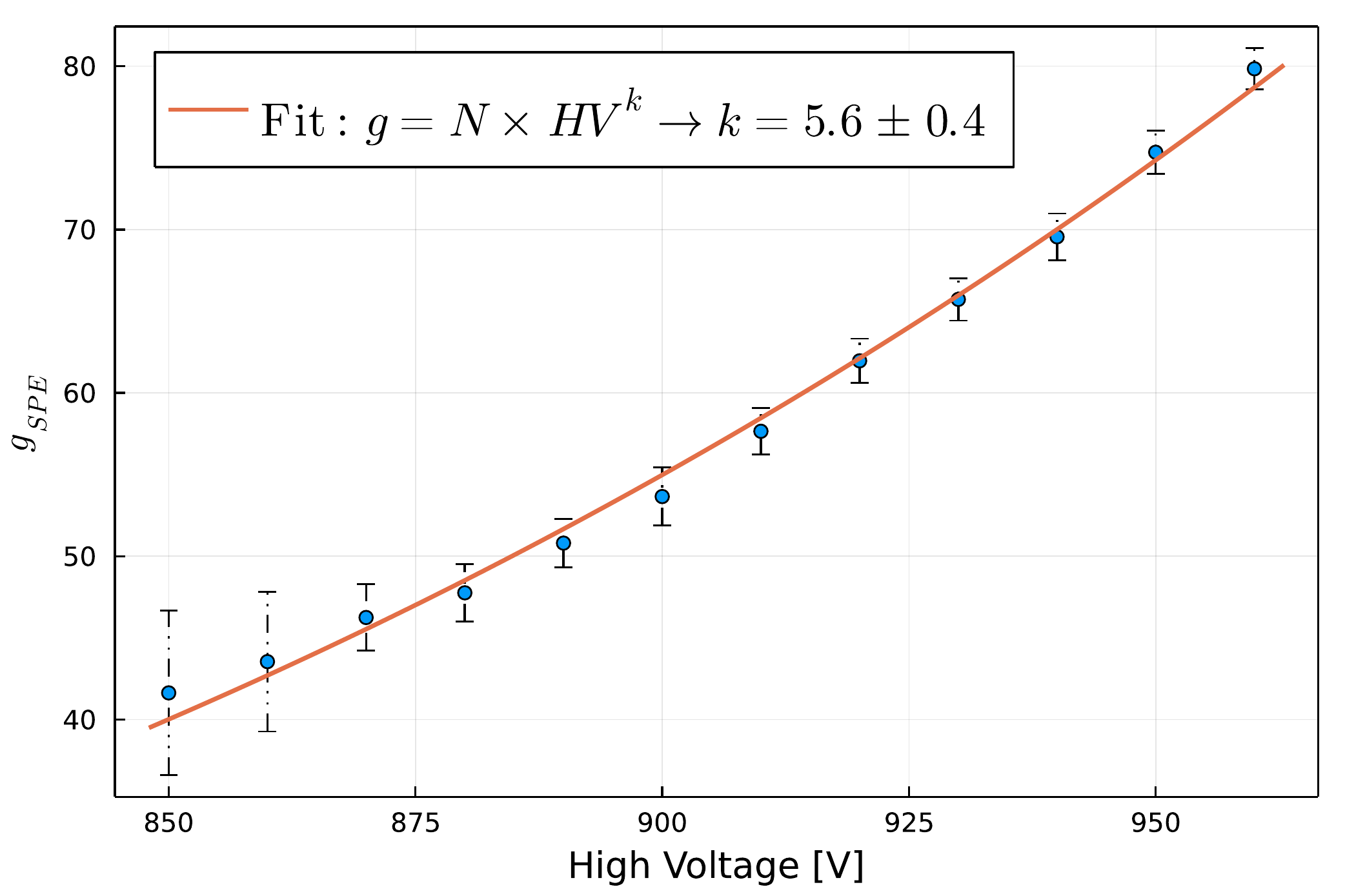}
    \caption{Left: Typical SPE distributions for different values of HV. Right:  Gain as function of HV.}
    \label{fig:pmt-calibration}
\end{figure}

A voltage scan, i.e. determination of the charge distribution as a function of PMT bias voltage, around the recommended operational high voltage of 900~V was also performed. Finally, the gain was defined as:
\begin{equation}\label{eq:gain}
g_{\text{SPE}}=\mu_{\text{1PE}}-\mu_{\text{Ped}}.
\end{equation}
with $g_{\text{SPE}} = \text{charge / SPE}$ being the charge per SPE. A typical SPE distribution is shown in Figure~\ref{fig:pmt-calibration}~(Left). The main components of Equation~\ref{eq:spe} (pedestal and 1st p.e.) can be clearly observed. Figure~\ref{fig:pmt-calibration}~(Right) shows the relation between the gain (SPE) and the HV, as expected this relation follows a potential law.

\subsection{Simulation of the setups}
Accurate simulations of the setups used for these studies were developed using the Geant4 simulation toolkit \cite{Agostinelli:2002hh}. A complete geometrical description of the trigger system described in section~\ref{section_trigger}, the $^{207}$Bi source and its container as well as a detailed description of the PMTs and scintillator samples were implemented.  

\begin{figure}[h]
    \centering
    \includegraphics[width=0.45\textwidth]{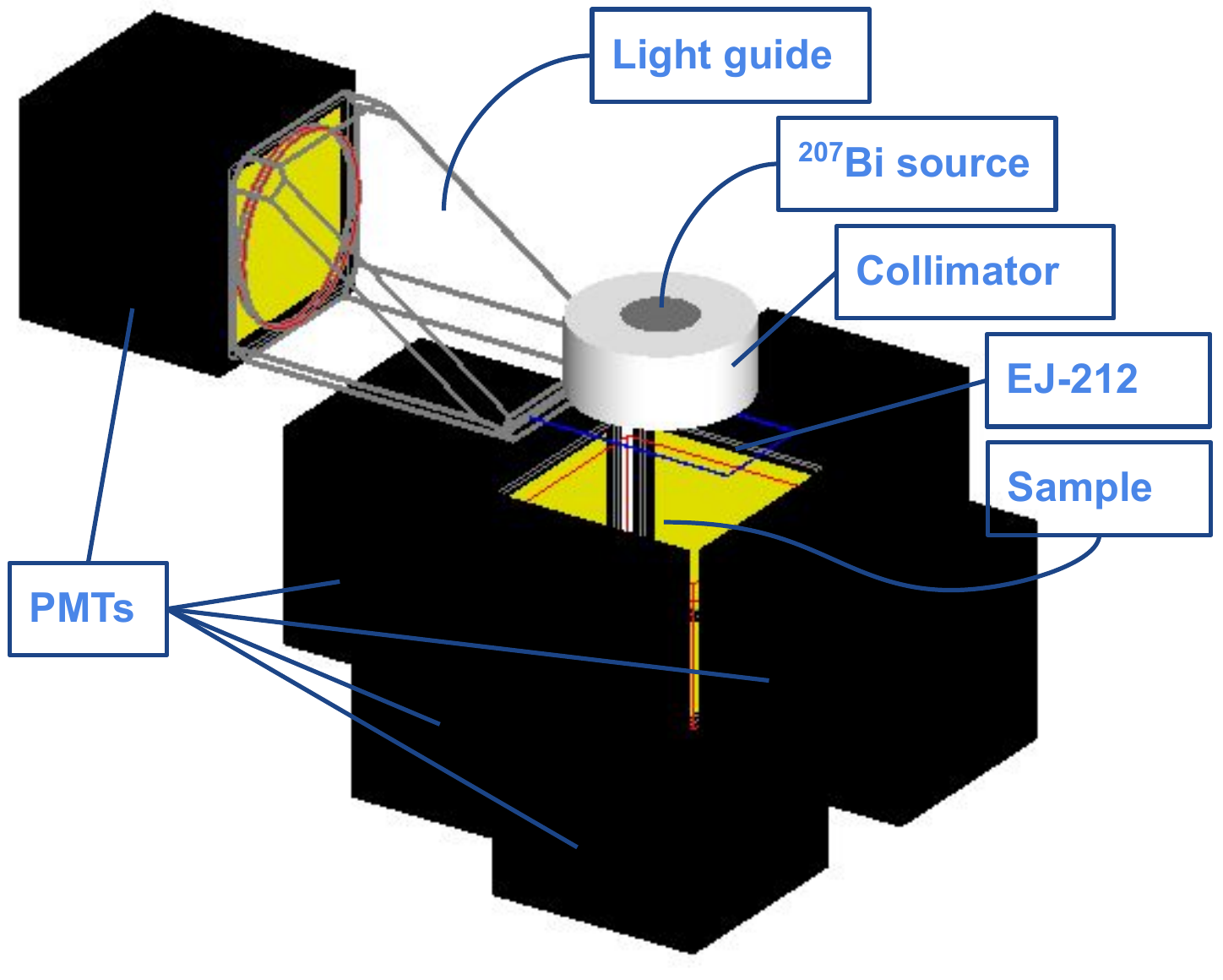}
    \includegraphics[width=0.45\textwidth ]{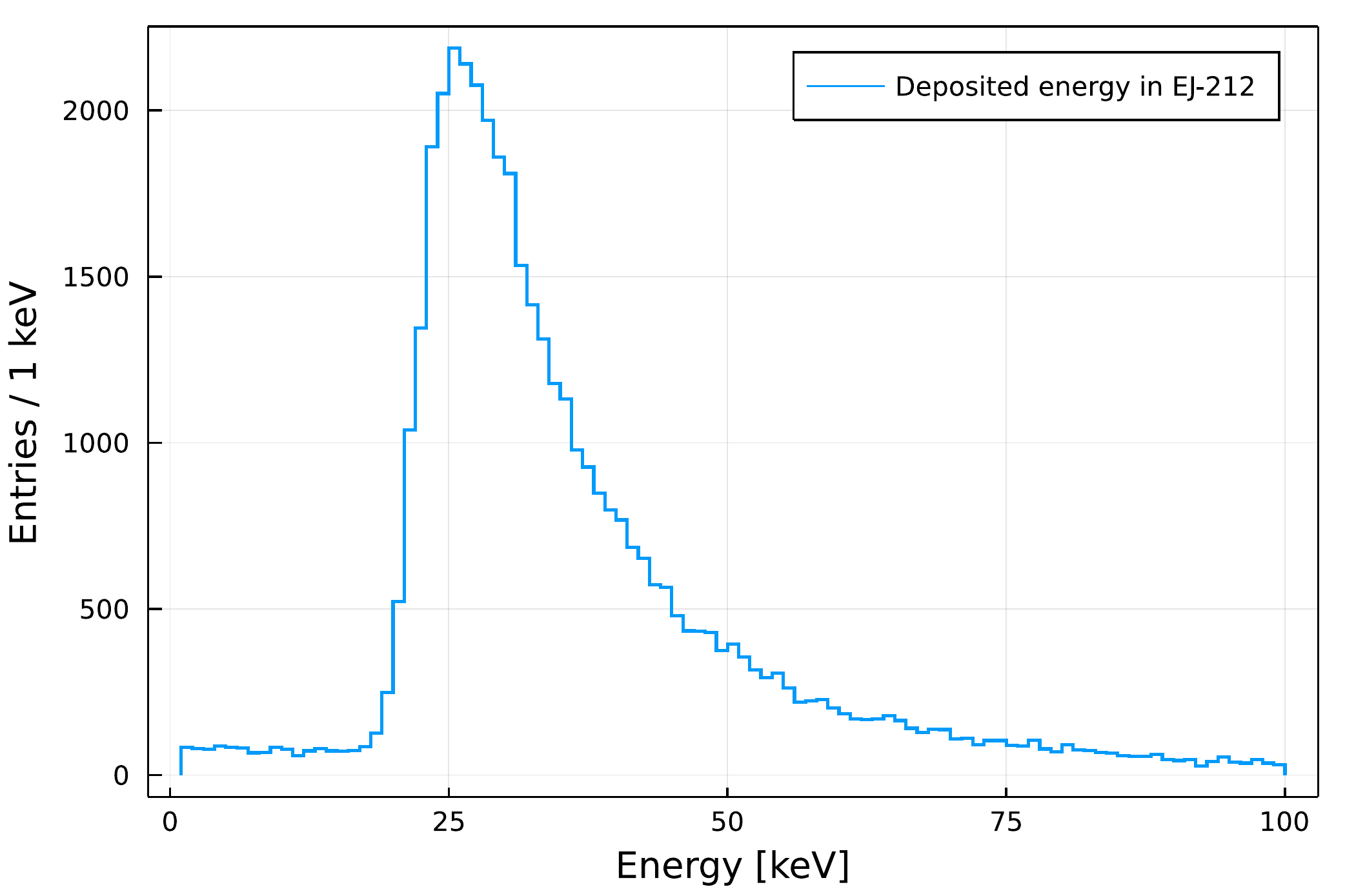}
    \caption{Left: Geant4 visualization of a setup to study the light output and light yield of scintillator samples.  Right: Monte Carlo true deposited energy by electrons from $^{207}$Bi source in the thin EJ-212 scintillator.}
    \label{fig:MC-Edep}
\end{figure}

\begin{figure}[h]
    \centering
    \includegraphics[width=0.7\textwidth ]{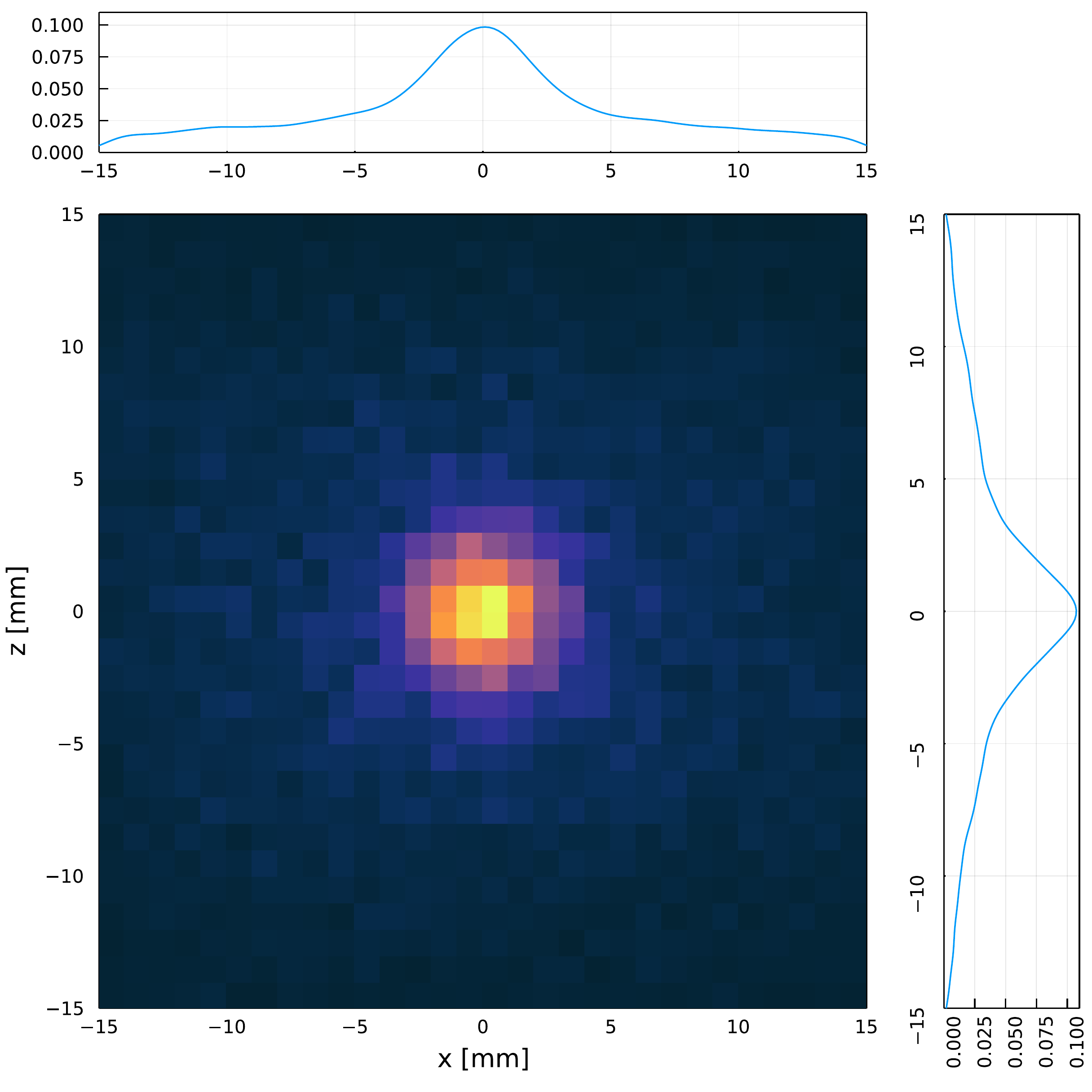}
    \caption{First position of interaction of triggered collimated electrons reaching the scintillator samples being measured. About 68~\% of the electrons are distributed in a circumference of 7~mm radius. The $^{207}$Bi source was placed just above the center of the sample ( coordinates [0,0] in the plane z-x).}
    \label{fig:MC-beamspot}
\end{figure}

The goal of these simulations was to determine the energy deposited by collimated electrons emitted by the  $^{207}$Bi source in the samples being measured. The electrons lose part of their energy in the source container and in the thin EJ-212 scintillator before reaching the target samples. The distribution of deposited energy of collimated electrons from the $^{207}$Bi source traversing the EJ-212 scintillator is shown in Figure~\ref{fig:MC-Edep}~(Right). The electrons deposit on average 25~keV in the EJ-212 scintillator. The interaction of electrons in the EJ-212 scintillator lead to scattering in the direction of these collimated electrons. This results in a beam spot in the scintillator samples focused below the center of the collimator. The distribution of the first position of interaction of collimated electrons, previously interacting in the EJ-212 scintillator and reaching the scintillator samples (surface of $30\times30$~mm$^2$) in the setup is shown in Figure~\ref{fig:MC-beamspot}. About 68~\% of the electrons interact in a circumference with a radius of about 7~mm. It was found that the electrons from the $^{207}$Bi source can be fully absorbed in 3~mm of PEN material. Therefore, 2 samples of 1.7~mm thickness were stacked above each other for most of the measurements. 

\begin{figure}[h]
    \centering
    \includegraphics[width=0.7\textwidth ]{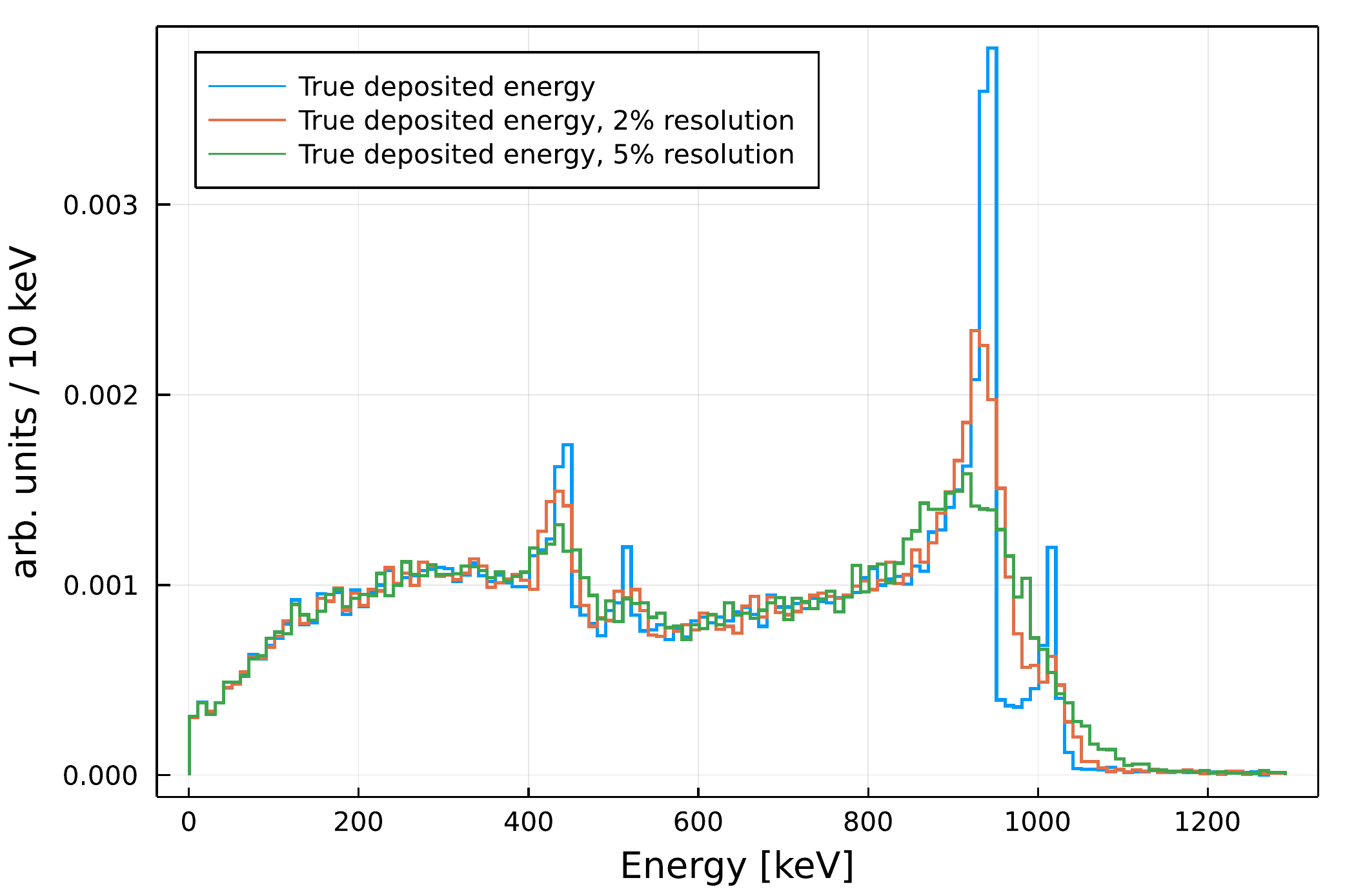}
    \caption{Monte Carlo true deposited energy by electrons from the $^{207}$Bi source. The blue line represents perfect energy resolution, while the orange and green lines represents an energy resolution of 2 and 5~\%, respectively.}
    \label{fig:MC-Edep-pen}
\end{figure}

The final deposited energy in the  target scintillator samples after energy losses in the source material and the EJ-212 trigger scintillator was computed and is shown in Figure~\ref{fig:MC-Edep-pen}. The four main electron lines can be observed in the Monte Carlo (MC) true deposited energy. However, taking into consideration the discrete energy resolution of a few percent, only two peaks are expected to be observed using a setup with PMTs. The two peaks have a mean energy of around 420 and 930~keV, respectively. The effect of a finite energy resolution is taken care of by convoluting the MC true energy spectrum with a Gaussian with width corresponding to the expected energy resolution of the used PMTs. This is shown in Figure~\ref{fig:MC-Edep-pen} (orange and green histograms). 

The Geant4 simulation framework was also used to study light propagation and collection. Hence, the optical properties of the different components were implemented. To this end, the UNIFIED model of Geant4 was used \cite{unified_model_citation,SurfacesGeant4}. This model describes the physics processes that optical photons undergo at a surface with respect to the surface properties. In this framework, parameters including the scintillation yield, emission spectrum, bulk absorption length, excitation time and surface roughness\footnote{In the Unified model a surface is modeled as consisting of micro facets. The angle $\alpha$ of a micro facet with respect to the average surface normal is sampled from a normal distribution. The  standard deviation  of  the  distribution  of  the  micro-facets  orientations is defined by sigma alpha $\sigma_{\alpha}$, which is required as input for the simulations. A higher $\sigma_{\alpha}$ implies higher roughness of the surfaces. On the other hand,  $\sigma_{\alpha}=0$ corresponds to a perfectly polished surface. The $\sigma_{\alpha}$ must be combined with the type of specular reflections to take effect \cite{unified_model_citation,SurfacesGeant4}.}
need to be defined. 
Some of these parameters such as excitation time and the emission spectrum can be measured directly. For the others, simulations or a calibrated scintillator sample as reference is required. In the next sections, results of direct measurements and estimation of these parameters are reported.     
\section{Surface characterization}
\label{sec:surface}
In the configuration of the \legend-$200$ and \legend-$1000$ experiments, most of the PEN holders are placed between two Ge detectors. An ideal holder should collect and propagate as much light as possible to the lateral sides where the optical fibers are located, in this way maximizing light detection efficiency. 
Hence, the surface quality of the PEN holders is of paramount importance.
Since the PEN holders were cut from a larger tile produced by injection molding, the larger surfaces facing the Ge detectors correspond to the molded sides, while the lateral sides correspond to a rough cut.
This method ensures having surfaces with excellent quality on the molded sides. 

\begin{figure}[h]
   \centering
   \includegraphics[width=0.45\textwidth]{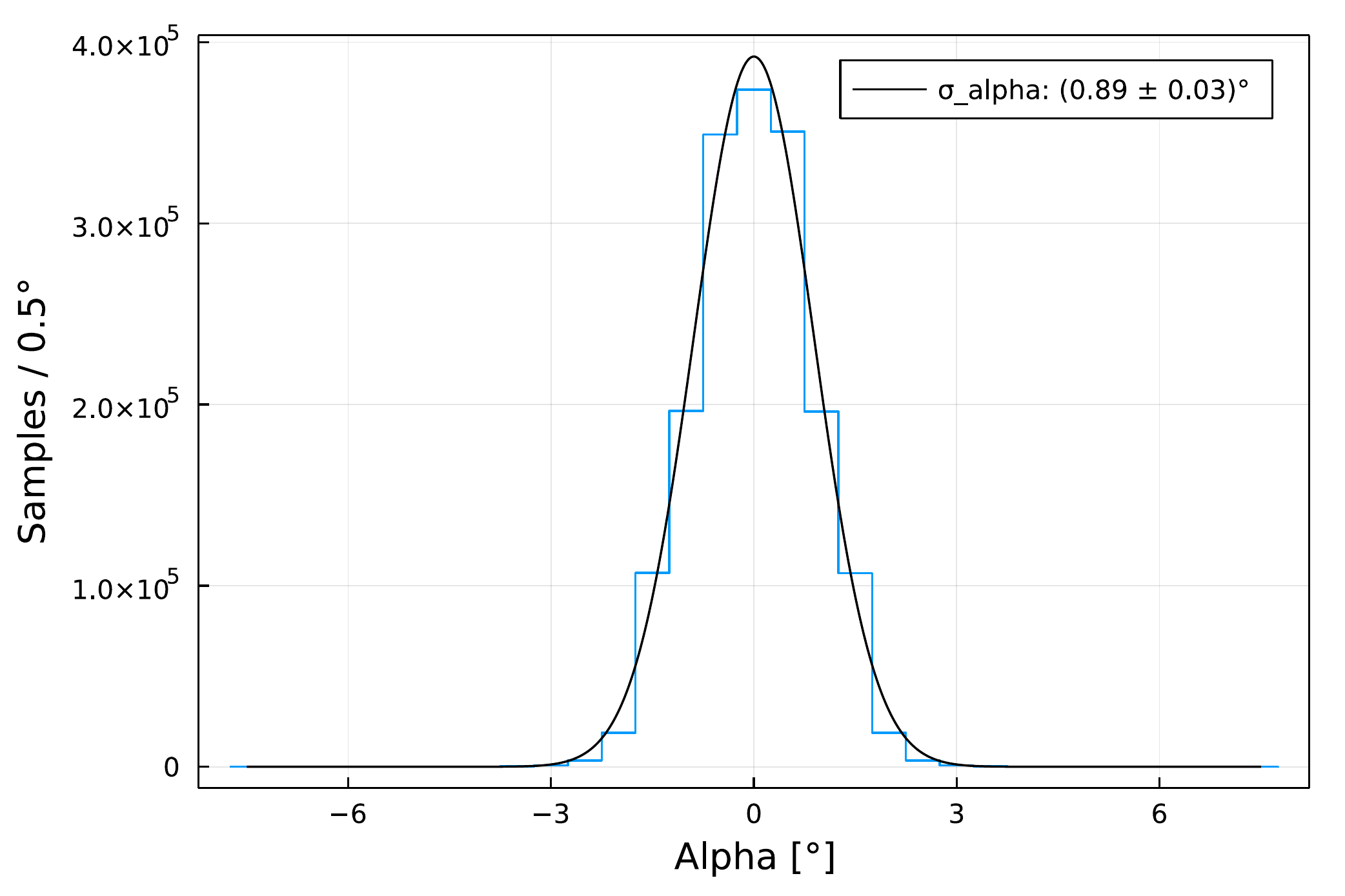}
   \includegraphics[width=0.45\textwidth]{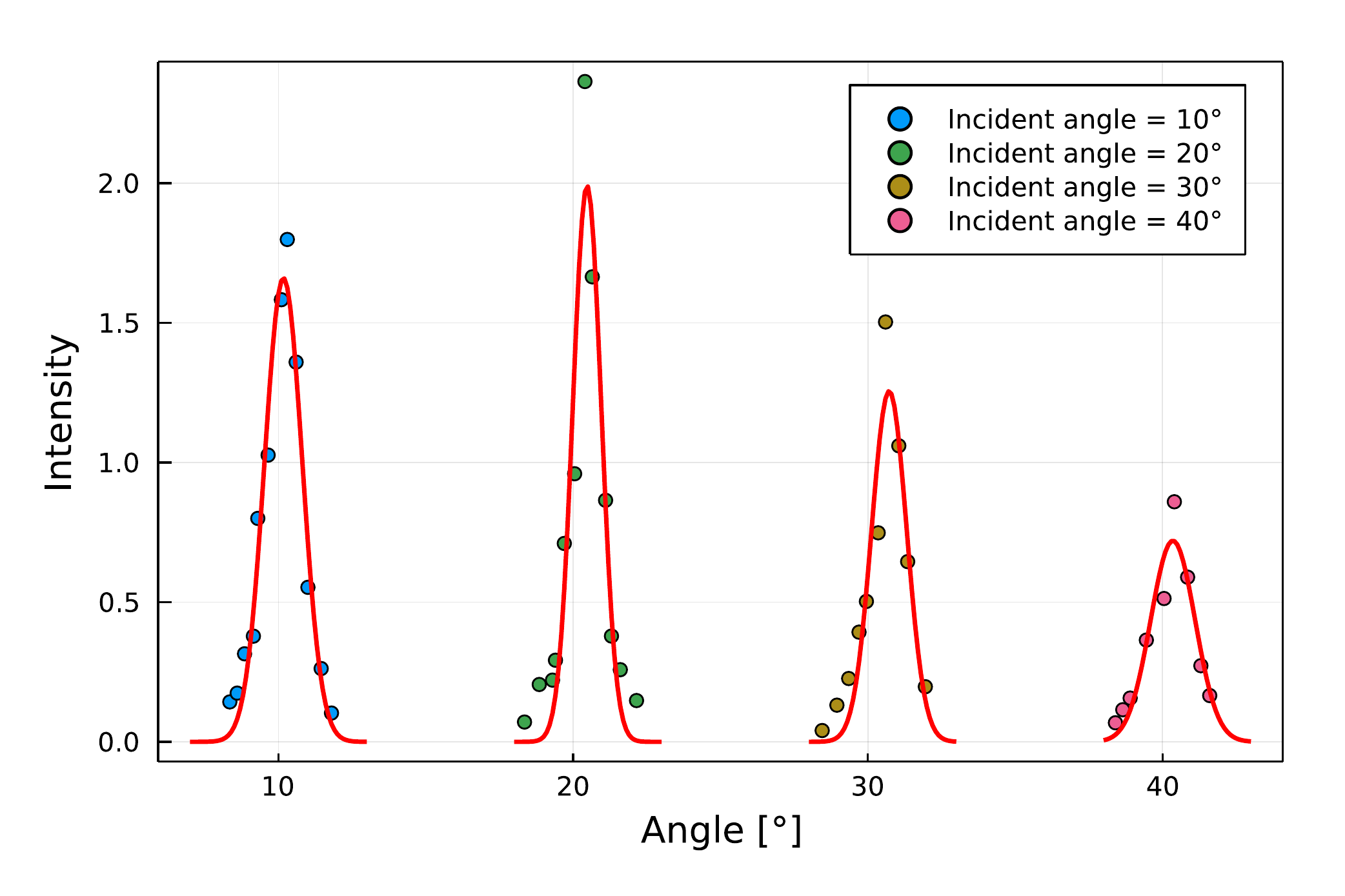}
   \caption{Left: Angular distribution of the slopes measured using a 3D image with micrometre resolution. Right: Intensity of  reflected light on the PEN surface for different incident angles.}
   \label{fig:ORNL_reflectivity} 
\end{figure}

The surface of the finished holders was characterized using a Keyence VHX-$6000$ digital microscope. This device allowed for a closer and detailed examination of the surfaces with a resolution of $\approx440$\,nm in each of the $x$ and $y$ directions and $\approx10$\,nm in $z$ for chunks of surface holder of about 1 mm$^2$. Using
the 3D profile of the surfaces thus obtained, a measurement of the surface roughness expressed as  $\sigma_{\alpha}$ was estimated. This $\sigma_{\alpha}$ is defined as the average of all the slopes calculated from the center of each point in the grid to all surrounding points (8 in total) individually. The distribution of the slopes computed in this way is shown in Figure \ref{fig:ORNL_reflectivity}. Using this distribution, a  $\sigma_{\alpha}$ of {$(0.89\pm0.03)^\circ$} for the molded sides was measured. Measurements with different samples revealed that in some cases (samples with scratches) a surface roughness up to $3^\circ$ can be present.

Additional characterization of the surface of the holders was performed using a special device developed to study how light incident at a given angle will result in reflected light intensity to vary as a function of angle. By fitting the measured power of the reflected light with a Gaussian distribution, it was possible to estimate $\sigma_\alpha$ as the sigma derived in the Gaussian fit. The measurements taken with this instrument used light at $\lambda = 513$\,nm.
The sensitivity of the device was determined and calibrated using a mirror placed on top of the PEN holder. These calibration results showed a limit on the sensitivity to the measurement of $\sigma_{\alpha}$ of $0.40^\circ$ for polished surfaces with this device. Once the setup was calibrated, a  scan using the PEN holder was performed. To avoid reflections from the back surface of the PEN holder, it was painted with matte black paint so the photons arriving at this surface would be absorbed. Taking the average of these measurements and using the accuracy determined with the mirror measurements, a $\sigma_{\alpha}$ defined as $\sigma_{\alpha}=\sqrt{\sigma^2_{\alpha}(\text{sample})-\sigma^2_{\alpha}(\text{Mirror})}$ of $(0.46\pm0.07)^\circ$ was determined. 
This result (average of 4 points) falls within the distribution of $\sigma_\alpha$ obtained from the distribution measured with the 3D image method described at the beginning of this section (thousands of points). 
All these results confirmed that PEN holders with a high quality surface have been achieved. 

\section{Scintillation properties of the \legend-$200$ PEN scintillator}
\label{S:scintillation}
First evidence of PEN scintillation was reported in 2011, where a light yield comparable to a BC-408 Polyvinyltoluene~(PVT) sample was reported~\cite{pennakamura}.
Furthermore, previous measurements reported a PEN emission spectrum matching the sensitive regime of standard photo-sensors, which is crucial to optimize light collection efficiency.
In addition, the attenuation length and surface quality will determine the size and geometry of the scintillator components that can be used in the experiment. The measurement of all these parameters are reported and discussed in the following subsections.

\subsection{Time response}
The time distribution of the scintillation response of PEN is needed to determine the integration time required to read out all the signals generated by the scintillator. This parameter in general also limits the applications in which PEN can be used.
To measure the time response, PEN samples of 30$\times$30$\times$1.7~mm$^{3}$ were coupled to 5 PMTs as shown in Figure~\ref{fig:MC-Edep}~(Left). The samples were excited with electrons from the $^{207}$Bi source and the recorded PMT signals were used to determine the PEN decay time.

\begin{figure}[h]
    \centering
    \includegraphics[width=0.45\textwidth]{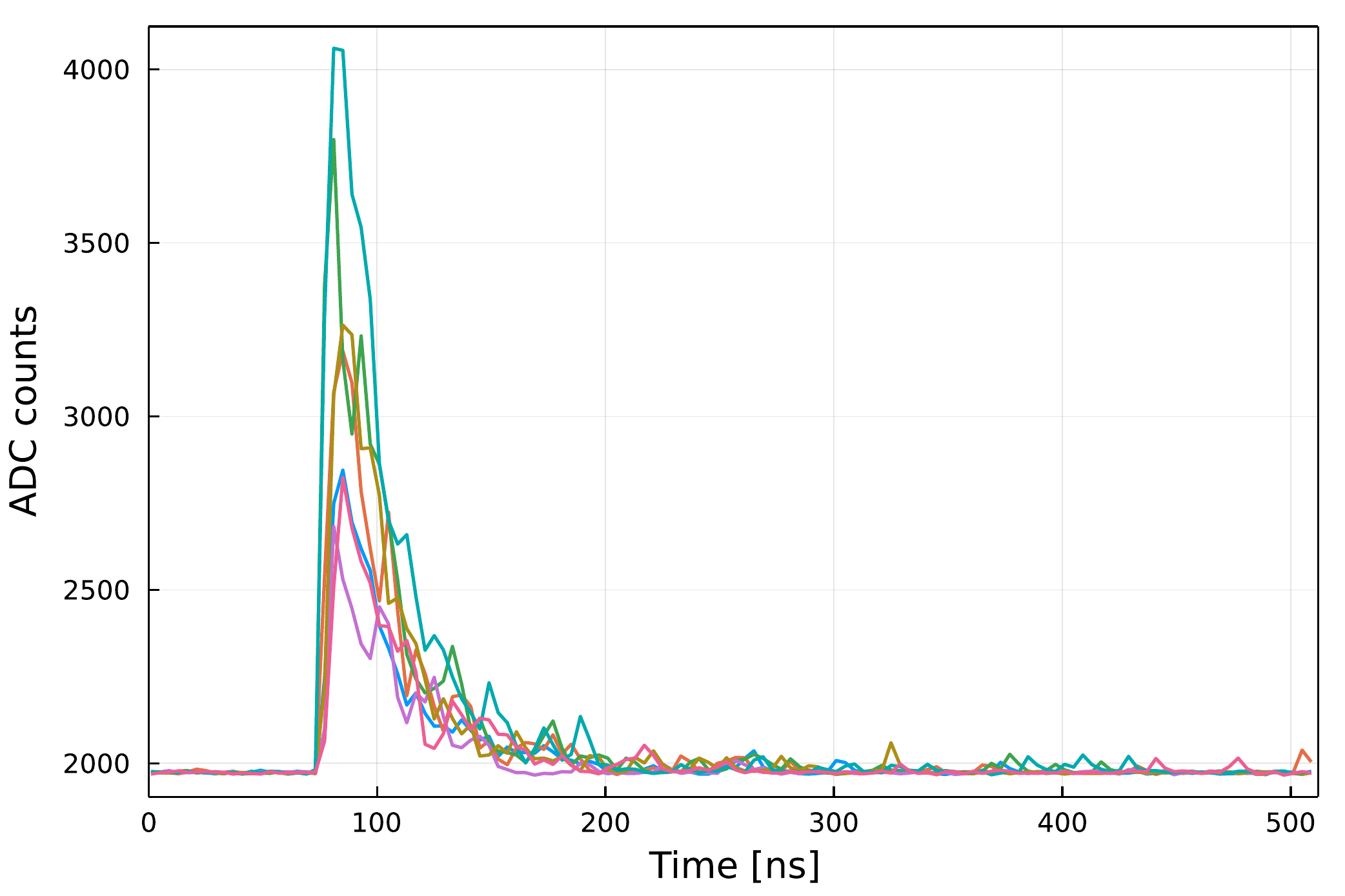}
    \includegraphics[width=0.45\textwidth ]{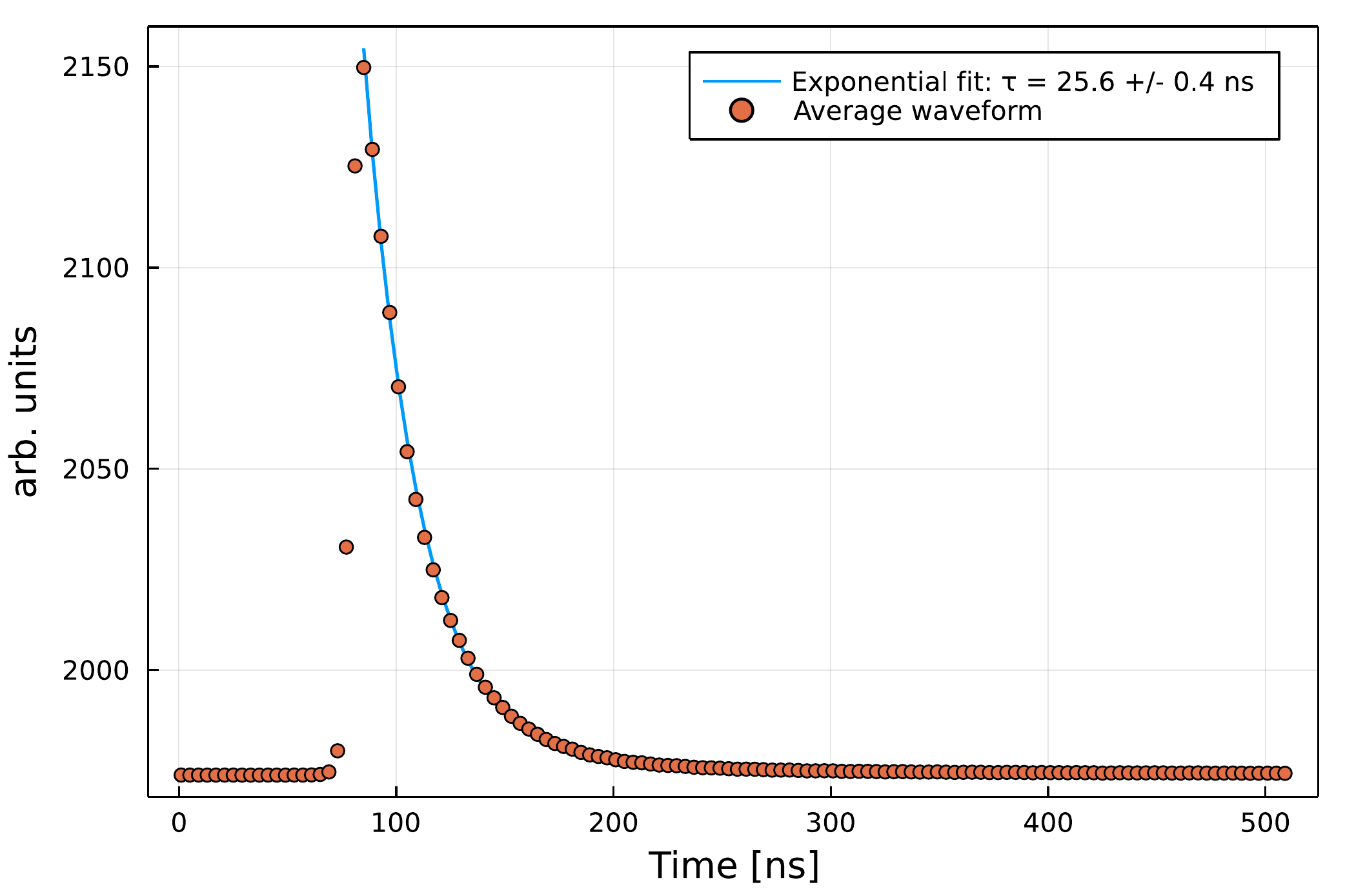}
    \caption{Left: Example of typical PEN waveforms sampled at 250 MHz readout using photo-multiplier tubes. The PEN samples are excited using electrons from a $^{207}$Bi source. Right: Average waveform fitted with an exponential function.}
    \label{fig:decay-time}
\end{figure}

Figure \ref{fig:decay-time}~(Left) shows typical waveforms of a PEN sample excited with electrons with energies around 1~MeV from the $^{207}$Bi source.
Taking advantage of the fast time response of the EJ-212 scintillator ($\sim 2$ns), it was used to align the PMT waveforms when coupled to the PEN sample. This alignment was performed by demanding the fast signal of the trigger PMT to always be at the same position in the waveform. Figure \ref{fig:decay-time}~(Right) shows the average waveform of all the signals selected in this way. The average waveform was then fitted using the exponential function 
\begin{equation}
f(t)= p_{0} + p_{1}e^{-t/\tau_{\text{PEN}}}
\end{equation}
with $\tau_{\text{PEN}}$ being the mean decay time or time constant of PEN. From data of 5 different PMTs and 4 PEN samples  a mean value of $\tau_{\text{PEN}}=(25.3\pm0.2)$~ns was found.
This value is in disagreement with previously reported values of 34.9~ns of PEN samples excited with 125~GeV protons \cite{Bilki:2019lep}, and slightly differs from \cite{Nakamura_2019} where $(28.9 \pm 0.2)$~ns was reported for PEN excited with beta particles from a $^{90}$Sr source. However, in the latter work two exponential functions were used for the fit.  Moreover, the difference can be also explained with a different composition of the material that can lead to different optical properties \cite{Marchi:2019moe}. In addition, it has been reported earlier that energy deposition in PEN by different particles leads to different pulse shapes. Thus, it is not unexpected that the scintillation response to protons will have a different time constant \cite{Efremenko:2019xbs,Hackett:2022xnk}.

\subsection{Emission spectrum}

\begin{figure}[h]
    \centering
    \includegraphics[width=0.7\textwidth]{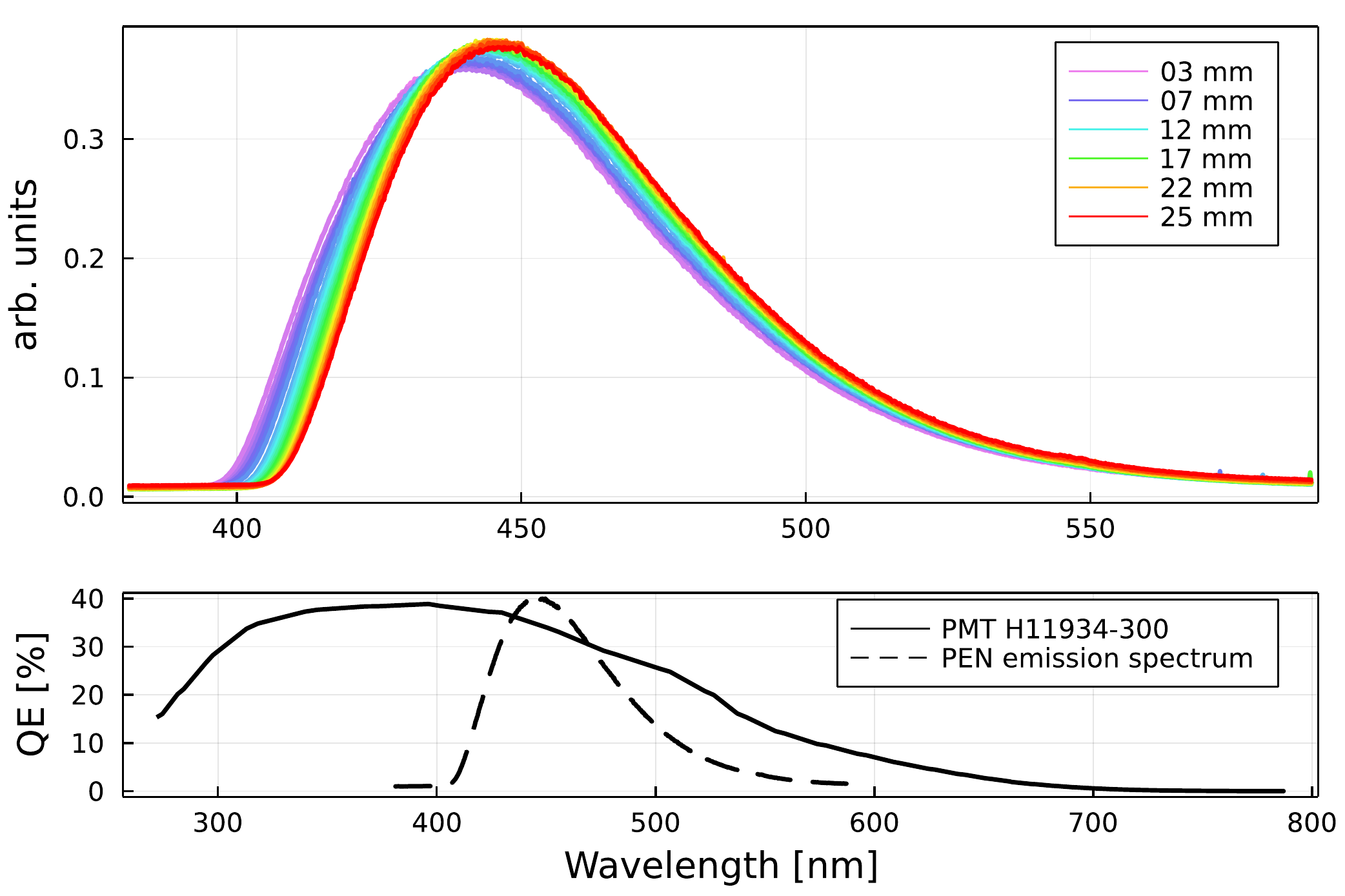}
    \caption{Top: Emission spectrum of a PEN sample excited with a collimated 340 nm UV LED at different distances with respect to the light collection point. Bottom: Quantum efficiency (QE) as a function of wavelength (solid line) for the H11934-300 PMT. The PEN emission spectrum (dashed line) is drawn for illustration. Quantum efficiency was taken from the PMT manual \cite{hamamatsuPMT}.}
    \label{fig:pen-spectrum}
\end{figure}

The PEN emission spectrum was studied using an Andor Shamrock 193i spectrograph, which  features a 0.2~nm wavelength resolution. PEN samples with 1.7$\times$20$\times$74~mm$^{3}$ dimensions, were  excited using a UV LED from Thorlabs (LED341W) with an emission wavelength of $(340 \pm 10)$~nm and a typical Full Width at Half Maximum (FWHM) of 15 nm. The PEN emission spectrum was measured by exciting the sample at different distances. In order to improve the position resolution of the excitation point with respect to the light collection window of the spectrometer, a light collimator allowing for a millimeter resolution was coupled to the LED. The results of these measurements are shown in Figure~\ref{fig:pen-spectrum} (top).     

It can clearly be seen that the peak of the emission spectrum shifts to higher wavelengths when the distance from the excitation point with respect to the collection window increases. In addition, this shift is higher at shorter wavelengths, which is a sign that the bulk absorption length is higher in this region. This self absorption behavior has been exploited to determine the bulk absorption length as a function of the photon wavelength as described in the next subsection.

In general, the PEN emission spectrum starts at around 390~nm, peaks around 440~nm and extends up to 600~nm. 
In this region, most of the photo-sensors can be used. As example, the quantum efficiency as a function of the wavelength can be seen in Figure~\ref{fig:pen-spectrum} (bottom) for the PMTs used in this work.

\subsection{Attenuation and bulk absorption length}\label{sec:pen_bulk}

\begin{figure}[h]
    \centering
    \includegraphics[width=0.7\textwidth]{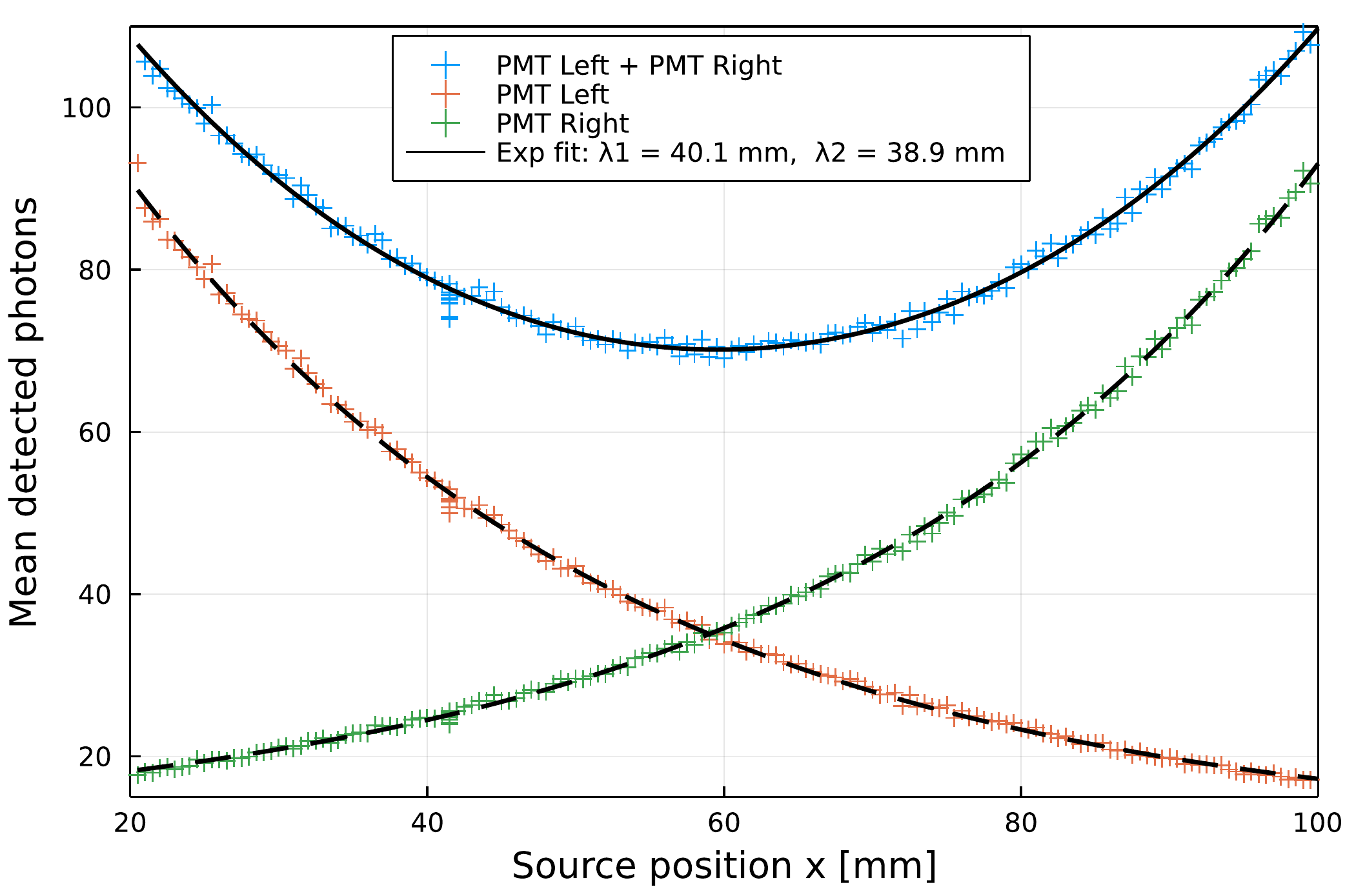}
    \caption{ Mean number of detected photons as a function of the distance from the excitation point to the PMTs. A long PEN sample (100$\times$20$\times$1.7 mm$^{3}$) coupled to two PMTs was used. The sample was excited using collimated electrons from a $^{207}$Bi source.}
    \label{fig:att_pmts}
\end{figure}

Another important characteristic of scintillators is the light attenuation and bulk absorption length. 
To determine the attenuation of the \legend-$200$ PEN scintillator, a sample of 100$\times$20$\times$1.7~mm$^{3}$ was used in conjunction with two PMTs coupled to the two ends of the sample (sides of 20$\times$1.7 mm$^{2}$). The sample was then excited with electrons from the $^{207}$Bi source using the trigger system described in subsection~\ref{section_trigger}. A first transverse scan over the side corresponding to the width of the sample (20~mm) was performed in order to identify the center of the sample on this axis ($y$-axis). Once the center on the $y$-axis  was determined, a scan was taken moving the source lengthwise ($x$-axis) from PMT1 (Left) to PMT2 (Right) in steps of 0.5 mm, with 60 s of data taking in each position.  For each position the mean number of detected photons in each PMT as well as the sum of both PMTs were computed and are shown in Figure~\ref{fig:att_pmts}. 
In this configuration the center of the sample corresponds to $x=60$~mm.

The change in the total number of detected photons (blue points in Figure \ref{fig:att_pmts}) as a function of the distance $x$ from the excitation point to the PMTs was fitted using an exponential function of type: 
\begin{equation}\label{eq:exp_pmts}
f(x)=p_{0}e^{-x/\lambda_1 }+p_{1}e^{x/\lambda_2 }
\end{equation}
Taking the average of $\lambda_1$ and $\lambda_2$, an attenuation value $\lambda=(39.5\pm1.9)$~mm was found.
The attenuation value $\lambda$ measured in this way is geometry dependent as it depends on the mean distance travelled by the photons, determined by the actual geometry. A more useful quantity, independent of the sample geometry is the bulk absorption length as a function of the photon wavelength $(wl)$.
The bulk absorption length $\lambda_b(wl)$, or the distance light of a certain wavelength can travel through a medium, before its intensity decays by $1/e $ is defined by: 

\begin{equation}
    \lambda_b (wl) = - \frac{L}{ln(T)}
\end{equation}
\begin{equation}
    T = \frac{I}{I_0}
\end{equation}
where $L$ is the length the light travels in the medium, $T$ is the transmission of the light, $I$ is the transmitted intensity and $I_0$ is the incident intensity.
The bulk absorption length was determined using three independent methods.

\begin{figure}[h]
    \centering
    \includegraphics[width=0.37\textwidth]{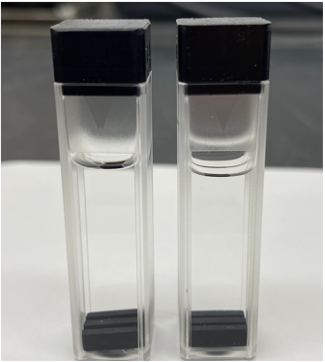}
    \includegraphics[width=0.55\textwidth]{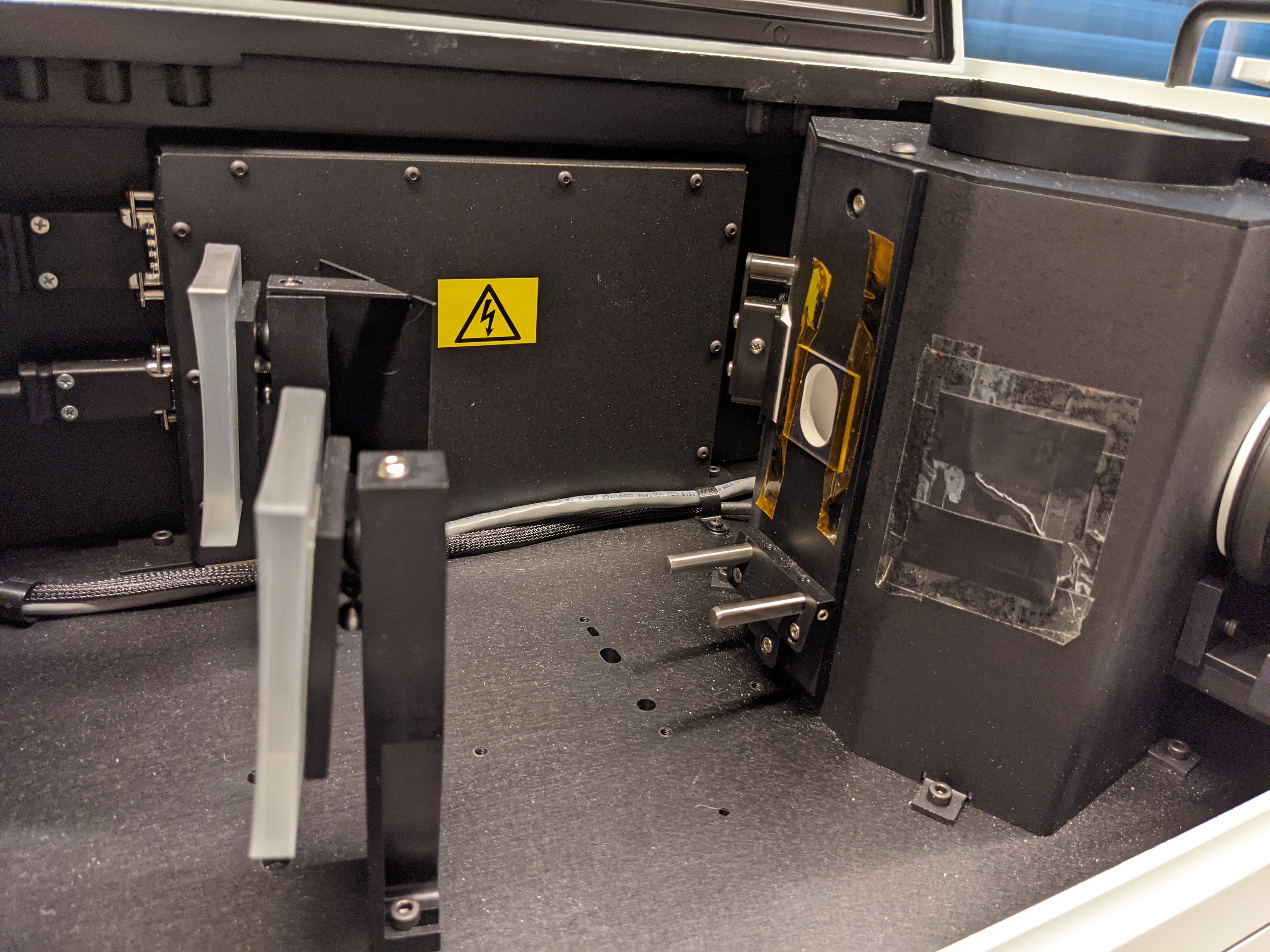}
    \caption{ Left: PEN pieces inserted into cuvette (on left) and reference cuvette (on right) without samples. Slotted black pieces were used in both the sample cuvette and the reference cuvette to ensure consistency. The cuvettes are then placed in a Shimadzu UV-2700i UV-Vis Spectrophotometer. Right: Inside of UV-VIS Lambda 850 spectrometer in transmission mode. A PEN sample was placed at the opening of an integrated sphere. }
    \label{fig:bulk_setups}
\end{figure}

First, $\lambda_b $ was measured using a Shimadzu UV-2700i UV-Vis Spectrophotometer. This instrument is a double beam, double monochromator spectrophotometer with a bandwidth of sub-nanometers and sensitivity from 200 to 800 nm. The double beam measures the transmission of both a sample and a reference, normalizing the intensity of the light source and improving the sensitivity of the instrument. 
A sample is commonly dissolved in a solvent and filled into a cuvette and the reference is an identical cuvette, filled with the solvent. 
As PEN cannot be easily dissolved, and dissolving the material could affect the true attenuation length of the bulk, an alternative method was developed to use this apparatus. 
PEN pieces were machined to fit perfectly into a BrandTech Macro UV-Cuvette and were fixed using a slotted black bottom and top fitting, designed to keep the face of the PEN plates parallel to the cuvette face, as well as to create a standard position as shown in Figure \ref{fig:bulk_setups} (Left). These slotted pieces were machined from black plastic, to ensure there would be no interference with the measurement and the reference liquid.

The cuvette was then filled with ethanol as the reference liquid. 
Ethanol was selected as it was chemically compatible with the cuvette material, the PEN plastic and the slot pieces. The ethanol used was purchased from Sigma Aldrich, HPLC Spectroscopic grade. 

A concerning factor on using plastic pieces submerged in a reference liquid is loss of light from changing refractive index. Without taking this loss of light into consideration, the attenuation length would have a systematic uncertainty and be lower. 
To predict this loss of light, optical simulations were conducted using the Geant4 package. 
In the simulation, it was assumed that the PEN, cuvette and ethanol had attenuation lengths much longer than the sample size. 
The factor, or fraction of light lost due to changing refractive index was determined by Equation  \ref{eq:factor}: 
\begin{equation}
    f_{PEN} = \frac{T_{\text{PEN~in~ethanol}}}{T_{\text{ethanol}}}
    \label{eq:factor}
\end{equation}

\begin{figure}[h]
    \centering
    \includegraphics[width=0.75\textwidth ]{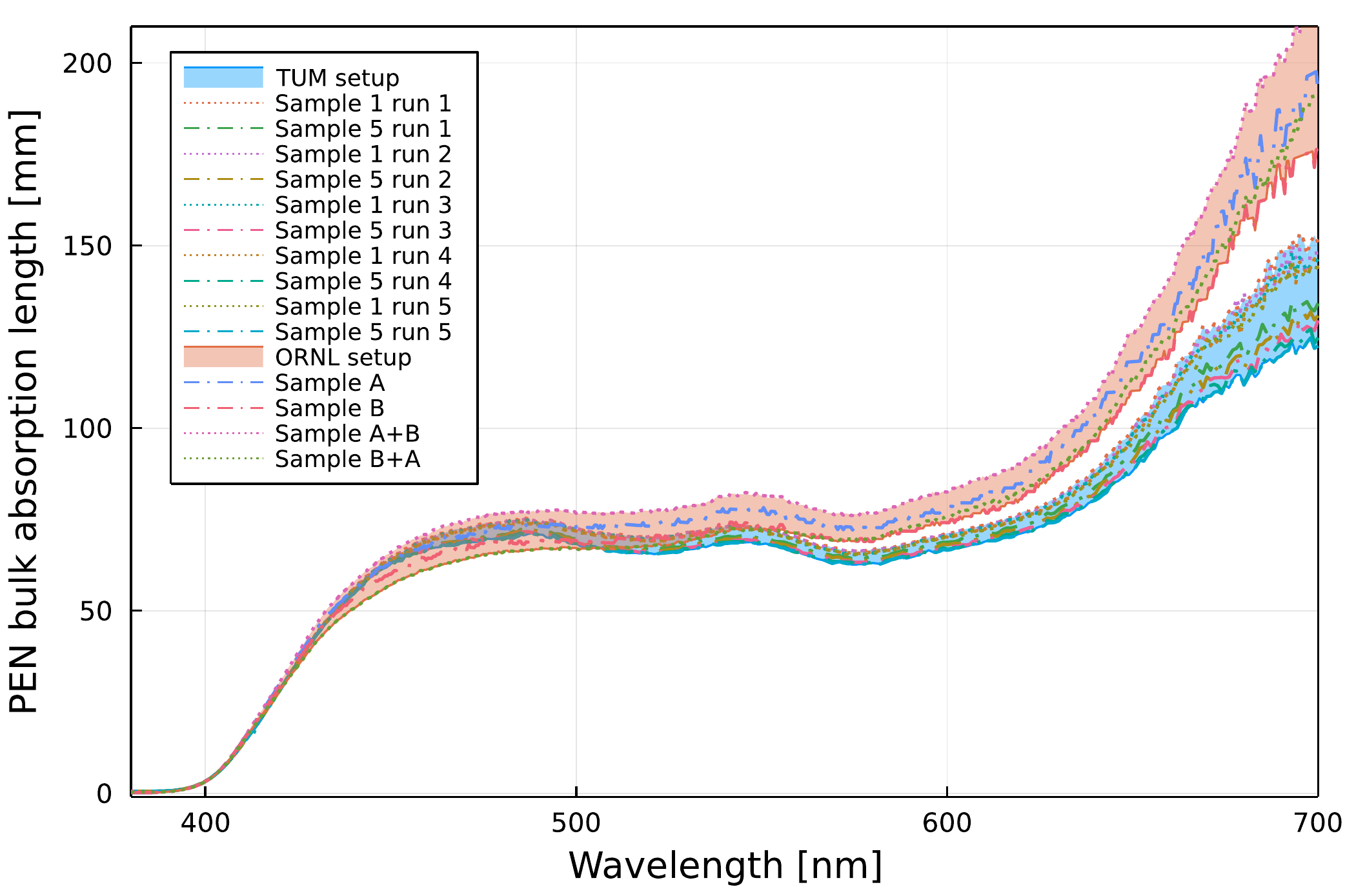}
    \caption{Measured bulk absorption length in PEN as a function of the photon wavelength using different methods. The color bands group the results of each setup. 
    In the region of the PEN emission spectrum the results are within each others uncertainty.
    }
    \label{fig:att_bulk}
\end{figure}

Where $T_{\text{PEN~in~ethanol}}$ and $T_{\text{ethanol}}$ correspond to the measured transmission at a given wavelength. The $f_{PEN}$ was fed into the attenuation length calculations to take this effect into account. 
To avoid surface effects, careful inspection was conducted using a microscope and bright lights.  Results obtained on the bulk absorption length from these measurements are shown in the orange band (ORNL setup) of Figure \ref{fig:att_bulk}.


The second direct method used a UV-VIS Lambda 850 spectrometer from Perkin Elmer. Using an integrated sphere allows the total amount of light, in both transmission and reflection mode, to be collected. The setup employs a monochromator allowing to select light of a given wavelength with sub-nanometer precision. 
Measuring both reflection and transmission allows to precisely estimate the amount of light that has been absorbed in the sample. 
Figure \ref{fig:bulk_setups}~(Right) shows a PEN sample placed in the setup before taking data in transmission mode. In reflection mode the sample is placed at the back of the sphere where a reference reflector is placed. Measuring the amount of reflected light allows to correct the amount of transmitted light  with respect to the incident light.
Two PEN samples of 30$\times$30$\times$1.7 mm$^3$ were used individually for these measurements. 
The setup was calibrated before each measurement and 5 runs with each sample were taken.
The results obtained from these measurements are shown in the blue band (TUM setup) of Figure \ref{fig:att_bulk}.

Finally, a third method using information of the PEN emission spectrum as shown in Figure~\ref{fig:pen-spectrum} was developed. This method exploits the information of the shift (to higher wavelengths) of the peak of the emission spectrum resulting from increasing absorption due to the increase of the distance between the excitation point and collection point.
However, this spectral information does not provide an absolute normalization factor. Therefore, using just this spectral information, only the shape of the bulk absorption length curve can be obtained.
To surmount this problem, the same sample was coupled to two PMTs.
The PMTs provided a robust method to determine the absolute normalization factor by measuring the mean number of detected photons according to the distance between the PMTs and the excitation point. Then, the PEN sample was excited at different distances with respect to the detection point as described at the beginning of this section. Later, this information was used to scale each spectrum of Figure~\ref{fig:pen-spectrum}.

The final ingredient required to compute the bulk absorption length is the mean total distance traveled by the optical photons before reaching the photo-sensors.
Since optical photons undergo multiple reflections on the surfaces of the material, the total path length is larger than the distance from the excitation point to the collection point. The total distance traveled by the photons between the excitation and the collection point was estimated using Geant4 optical simulations. Moreover, the path followed by the photons also depends on the surface roughness. The surface roughness will imply that the photons undergo more or less reflections before reaching the PMTs. As shown in section \ref{sec:surface}, high quality surfaces have been obtained.
Thus, simulations using a $\sigma_{\alpha}$ with values from 0 to 5 degrees for the molded sides were performed. These values correspond to polished and close to polished surfaces. On the other hand, for the lateral sides, simulations with values of  $\sigma_{\alpha}$ from 0 to 10 degrees were carried out. It was found that variation of the roughness of the surfaces within the range 0 to 5 degrees has only a minor impact on the total distance travelled by the photons.

\begin{figure}[h]
    \centering
    \includegraphics[width=0.7\textwidth]{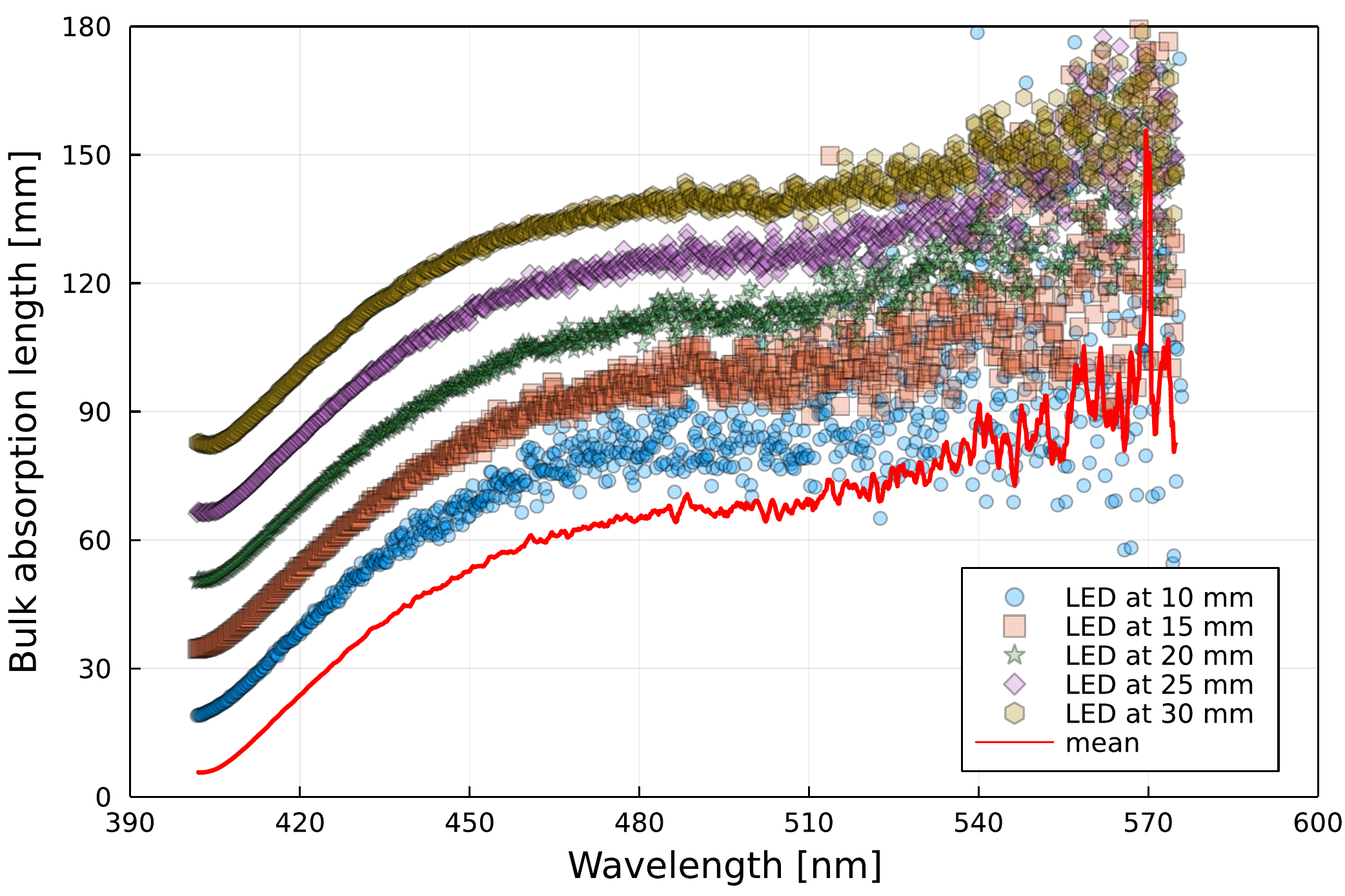}
    \caption{Estimated bulk absorption length in PEN as a function of the photon-wavelength using the spectral information of a sample excited at different distances from the collection point. A 15 mm shift on the y axis from curve to curve was used for visualization purposes.}
    \label{fig:att_bulk_iteration}
\end{figure}

All these ingredients were combined and using an iterative process, a bulk absorption length curve was obtained. 
For the first iteration an initial bulk absorption length of 150~mm was assumed for all wavelengths.
For each iteration the detected spectrum measured exciting the sample at position $x$ was scaled according to equation \ref{eq:exp_pmts} taking as reference the spectrum at $x = 5$~mm, which corresponds to the shortest distance between  collection point and excitation point. For each excitation position $x$ the corresponding bulk absorption length curve was found. Taking the average of all positions, a new bulk absorption length curve was obtained. This new bulk absorption curve was used in the simulations for the next iteration.  
After three iterations the results converged to a final bulk absorption curve. 
Figure~\ref{fig:att_bulk_iteration} shows the bulk absorption curve after 3 iterations for each position analyzed as well as the average of all positions to be used in the next iteration. 

\begin{figure}[h]
    \centering
    \includegraphics[width=0.7\textwidth ]{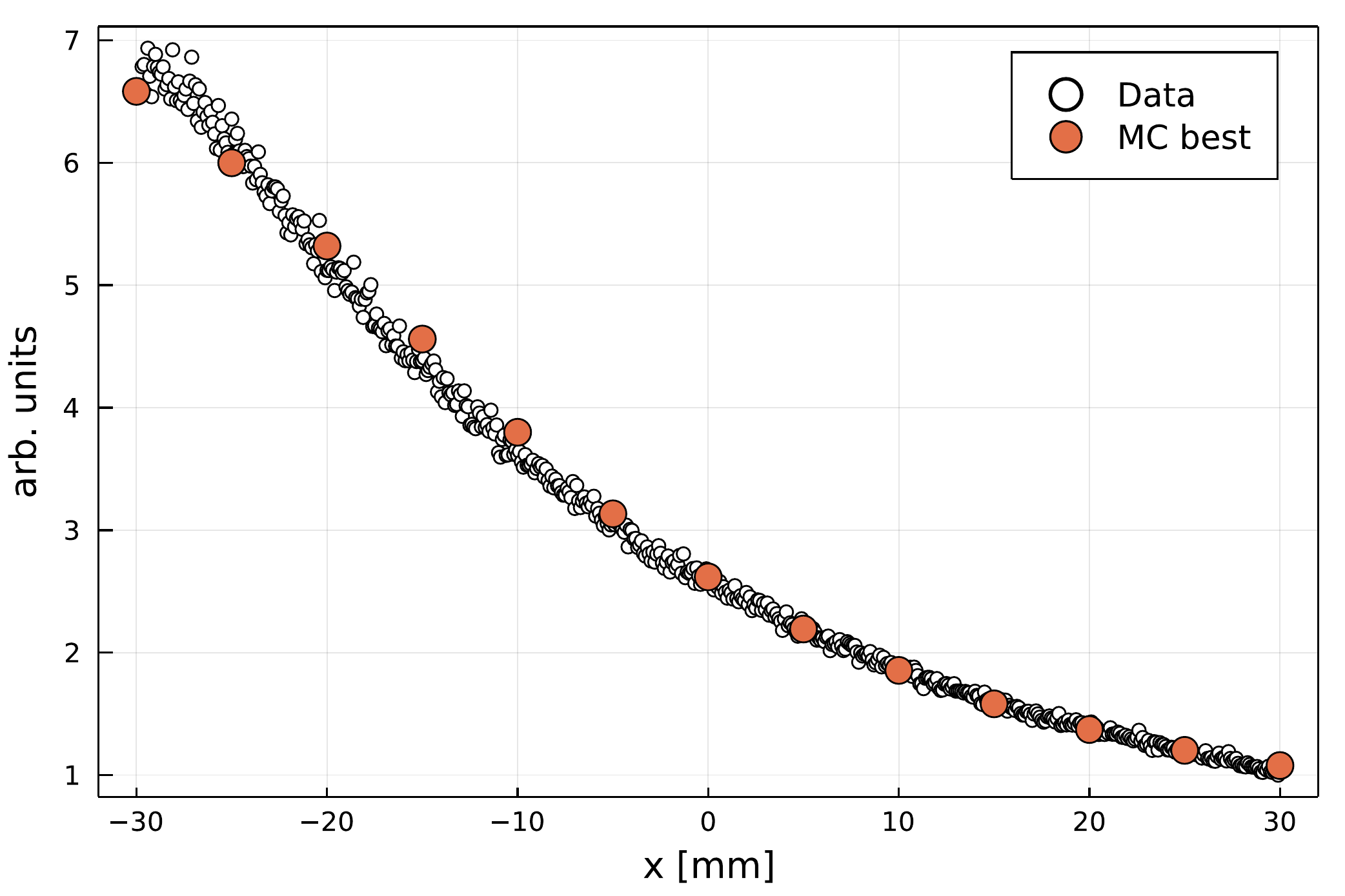}
    \caption{Data - Monte~Carlo comparison for tuned values of the bulk absorption length curve and surface roughness. A PEN sample of sample of dimension 70$\times$10$\times$1.7 mm$^{3}$ coupled to two PMTs was excited at different positions.}
    \label{fig:att_pmts_tunned}
\end{figure}

The results of the three methods reported in this section  are in good agreement. At the peak wavelength of the PEN emission spectrum around 440~nm the obtained bulk absorption length was 54.3$\pm$1.0~mm, 55.3$\pm$1.0~mm and 52.8$\pm$2.0~mm for the three methods, double beam with reference sample, integrated sphere with transmission / reflection measurement and iterative method, respectively. The third method gives a slightly lower value than the two first methods. However, the uncertainties are  higher because of  assumptions made for this method. 
The results were validated by using them as input for the simulation of the PMT data obtained with the measurement described at the beginning of this section. The simulations show an excellent agreement with the data for a long PEN sample as can be observed in Figure~\ref{fig:att_pmts_tunned}, thus giving confidence in the obtained results.

\subsection{Light output and absolute light yield}
The scintillation efficiency,
referred to as light yield, corresponds to the number of generated optical photons per amount of deposited energy for a given particle species. With known light yield, bulk absorption length, surface roughness and quantum efficiency of the photo-sensors, the light output (photons being detected by the PMTs) from a sample with a given geometry can be obtained. A higher light yield will result in a higher number of detected photons, translating into a higher detection efficiency and better background rejection efficiency. 

\begin{figure}[h]
    \centering
    \includegraphics[width=0.7\textwidth]{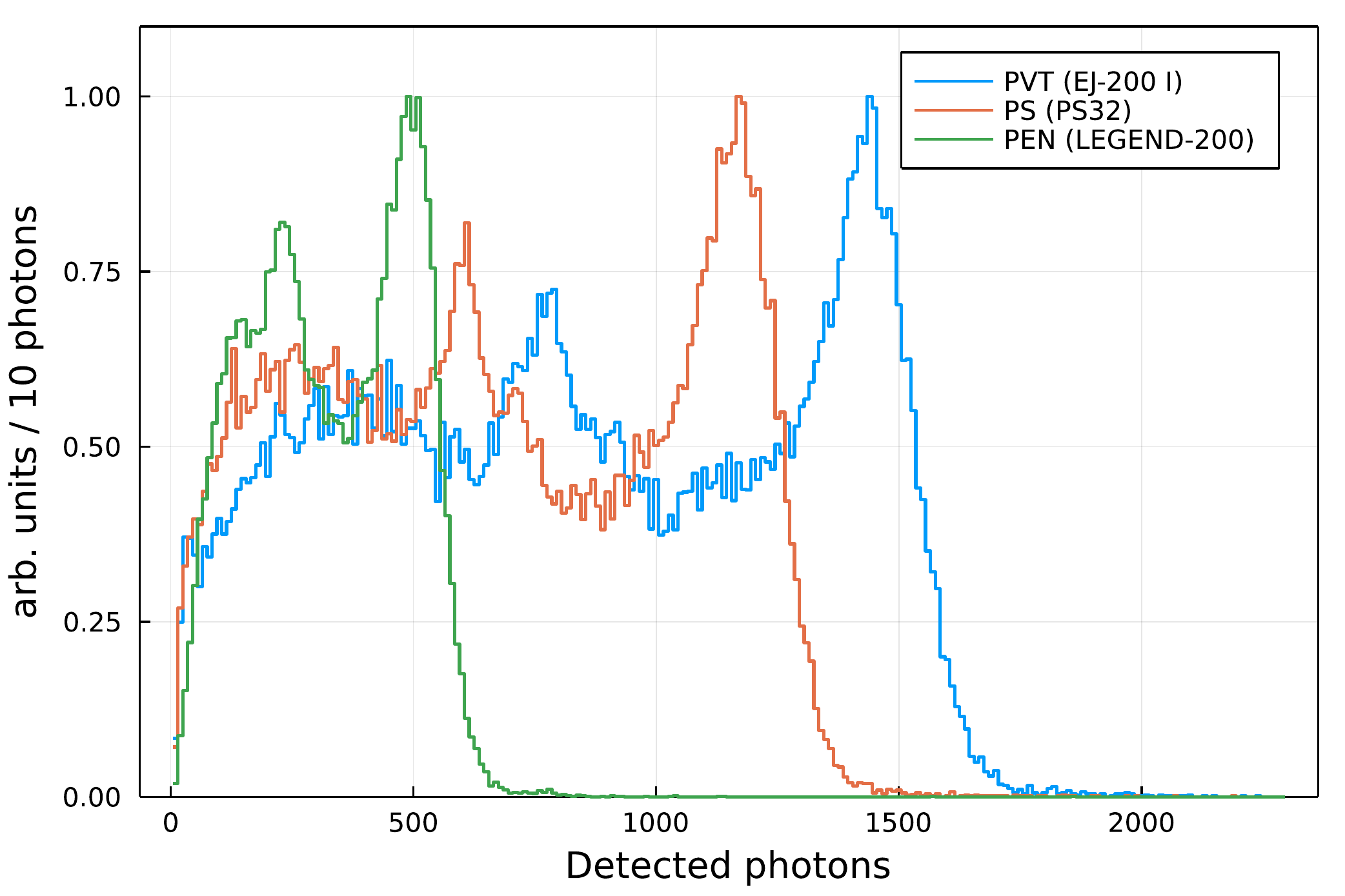}
    \caption{Number of detected photons for different scintillator samples excited with  $^{207}$Bi.}
    \label{fig:light-output-pen-pvt-ps}
\end{figure}

In order to measure the light output of PEN, square samples of 30$\times$30$\times$1.7 mm$^{3}$ were stacked and coupled to five PMTs, four lateral and one at the bottom as shown in Figure~\ref{fig:MC-Edep}~(Left). The samples were then excited with collimated electrons from the $^{207}$Bi source that was placed over the center of the samples. Scintillators with well known optical properties were procured and  samples of the same dimensions as PEN were machined. 
These samples allowed the detection efficiency of the setup to be calibrated and to have a direct comparison of the light output among the different scintillators.
To this end, EJ-200 and PS32 samples were used. These scintillators are based on polyvinyltoluene (PVT) and polystyrene (PS) respectively. The light yield of the EJ-200 and PS32 scintillators is reported to be 10000  and 8750 photons/MeVe$^{-}$, respectively.
The number of detected photons when using the $^{207}$Bi source to excite the three scintillator samples is shown in Figure~\ref{fig:light-output-pen-pvt-ps}.
The region around the peaks of each spectrum was fitted with a Gaussian distribution and this value was compared to the peaks of the deposited energy spectrum ($\sim$420~keV and $\sim$930~keV) obtained from simulations (see Figure~\ref{fig:MC-Edep-pen}) in order to obtain the light output.
As expected, the EJ-200 samples show the highest light output, with about 1500 photons/MeVe$^{-}$. 
A direct comparison of PEN with these two scintillators provided a lower limit of a light yield of 3000 photons/MeVe$^{-}$. However, while the bulk absorption of the EJ-200 and PS32 scintillators is of the order of meters, resulting in negligible attenuation effects  for these small samples, in the case of PEN this effect has to be taken into account.  
Thus, to determine the light yield of PEN, Geant4 optical simulations are needed, taking into account the effects of bulk absorption length and surface roughness using the measurements described in previous sections.

\begin{figure}[h]
    \centering
    \includegraphics[width=0.45\textwidth]{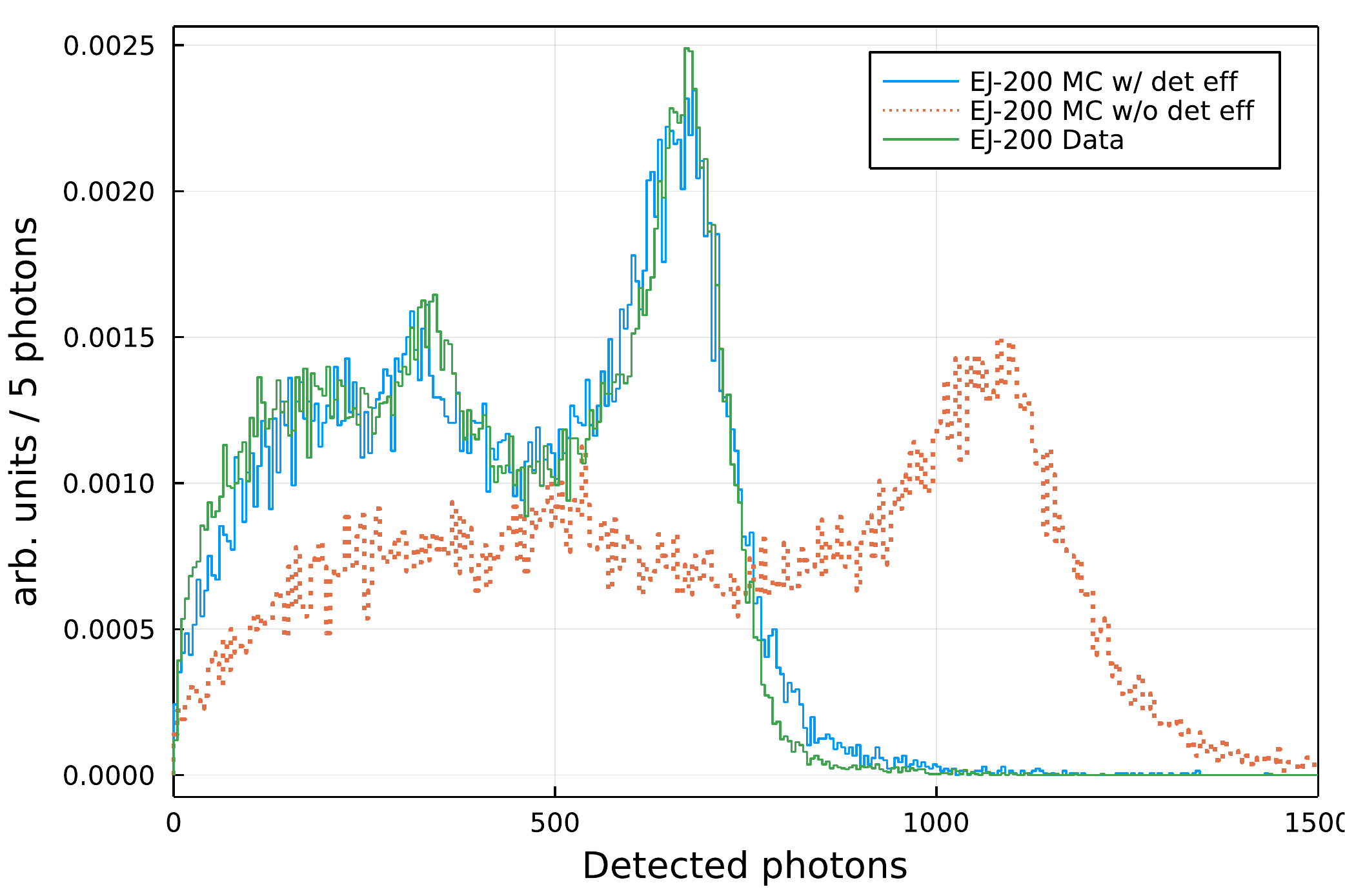}
    \includegraphics[width=0.45\textwidth ]{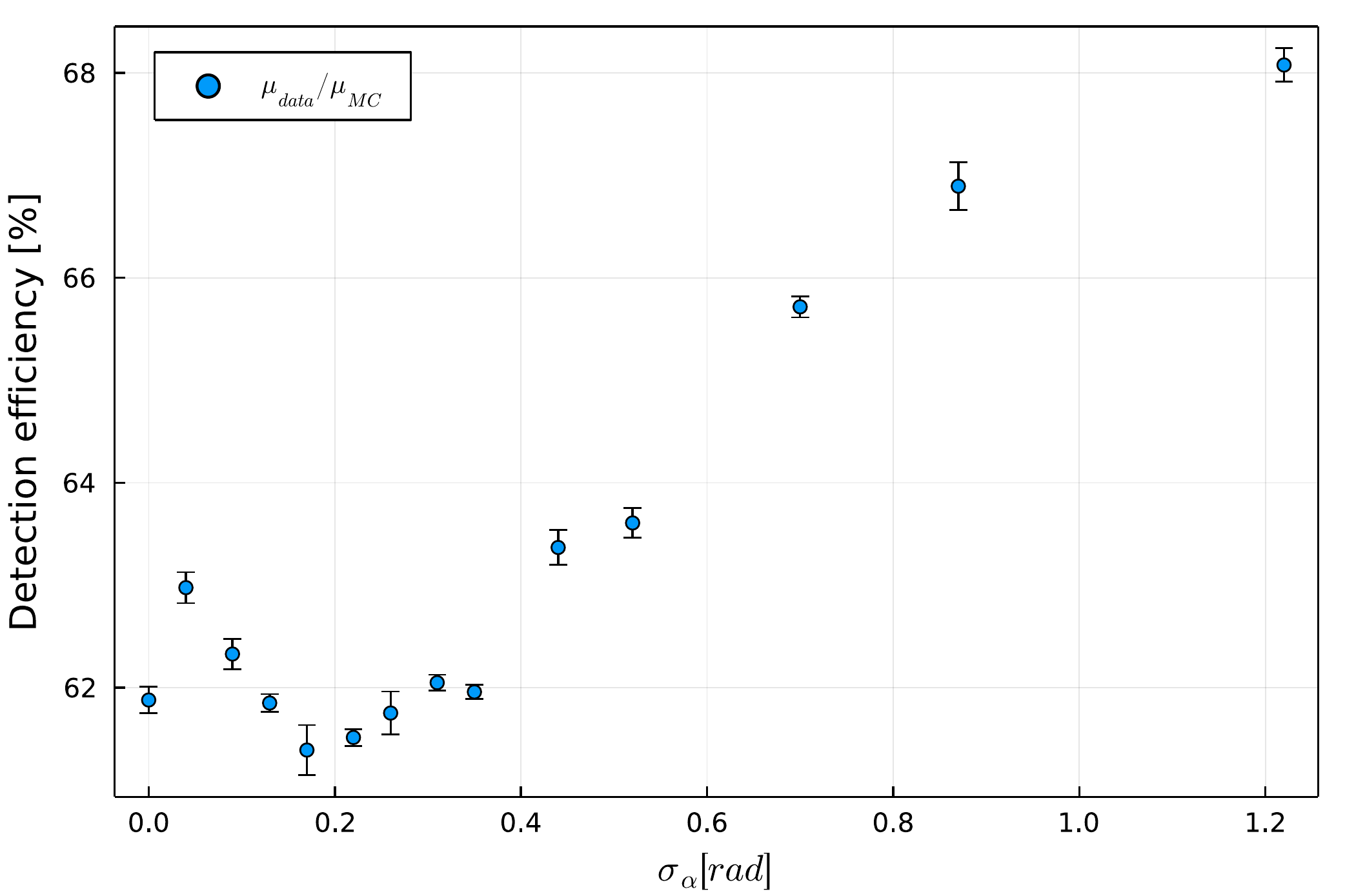}
    \caption{Left: Light output of EJ-200 samples excited with $^{207}$Bi source. The green histogram corresponds to data, while the blue and orange histograms correspond to the simulation assuming 100\% and 62\% detection efficiency respectively.  Right: Shows the detection efficiency from simulation as a function of the surface roughness for the EJ-200 samples.}
    \label{fig:mc-ej200-surfaces}
\end{figure}

First, taking advantage of the well known properties of the EJ-200 scintillator it was used to calibrate the detection efficiency of the setup. 
For this scintillator the bulk absorption length at 450~nm is about 380 cm, while the decay time and index of refraction are 2.1~ns and 1.58 respectively.  
The uncertainty on the light yield (10000 photons/MeVe$^{-}$) is about 2\% \footnote{From direct communication with ELJEN technology researchers, the 2\% uncertainty is equivalent to $\pm$~200 photons}, and it is included in the estimations of the systematic uncertainties of the PEN light yield.
The detection efficiency of the setup was determined by comparing simulations with data. 
Figure~\ref{fig:mc-ej200-surfaces}~(Left) shows the number of detected photons for the EJ-200 samples used as reference for the calibration of the detection efficiency of the setup. The simulated number of detected photons (orange) is higher with respect to the data (green). The region around the $\sim$1 MeV peak was fitted in both data and simulations and the values of the peaks thus obtained were compared to calculate the detection efficiency of the setup. Once the simulations were corrected for the detection efficiency of $\sim$62\% (blue histogram) an excellent agreement between data and simulations was achieved.

An unknown parameter of the  EJ-200 samples is the surface roughness. To estimate the uncertainty originating from this parameter, 3D images were taken and estimations of $\sigma_{\alpha}<0.2\text{ rad}$ 
were obtained following the same procedure as described in section \ref{sec:surface}. Then, simulations with different surface roughness were carried out. Figure~\ref{fig:mc-ej200-surfaces} (Right), shows the impact of the surface roughness for the EJ-200 samples on the determination of the detection efficiency of the setup. In the range of the experimentally obtained $\sigma_\alpha$,   an uncertainty of 2\% can be inferred. This effect is small because of the small size of the samples and the long bulk absorption length of the EJ-200 scintillator.

\begin{figure}[h]
    \centering
    \includegraphics[width=0.45\textwidth ]{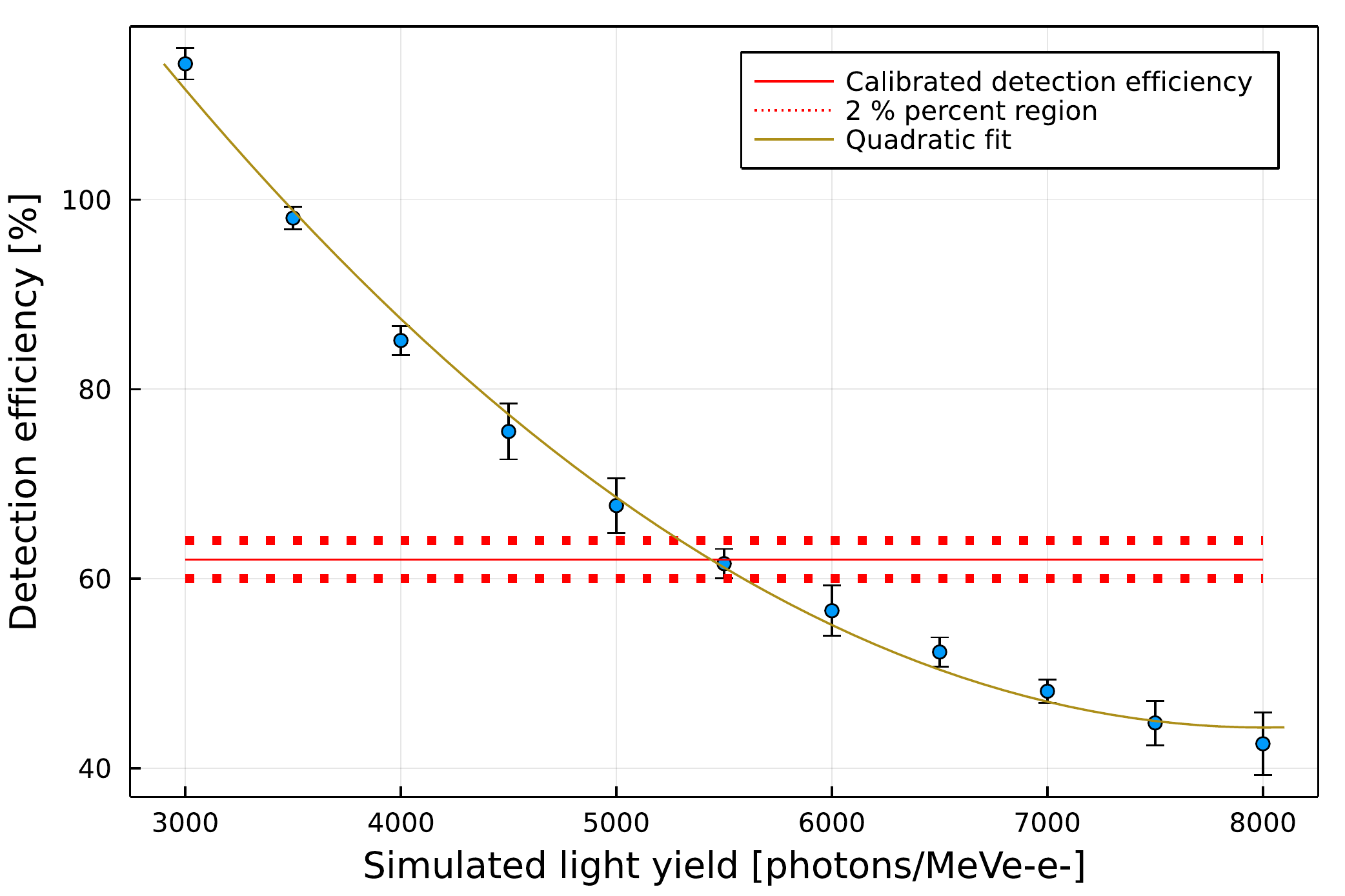}
    \includegraphics[width=0.45\textwidth]{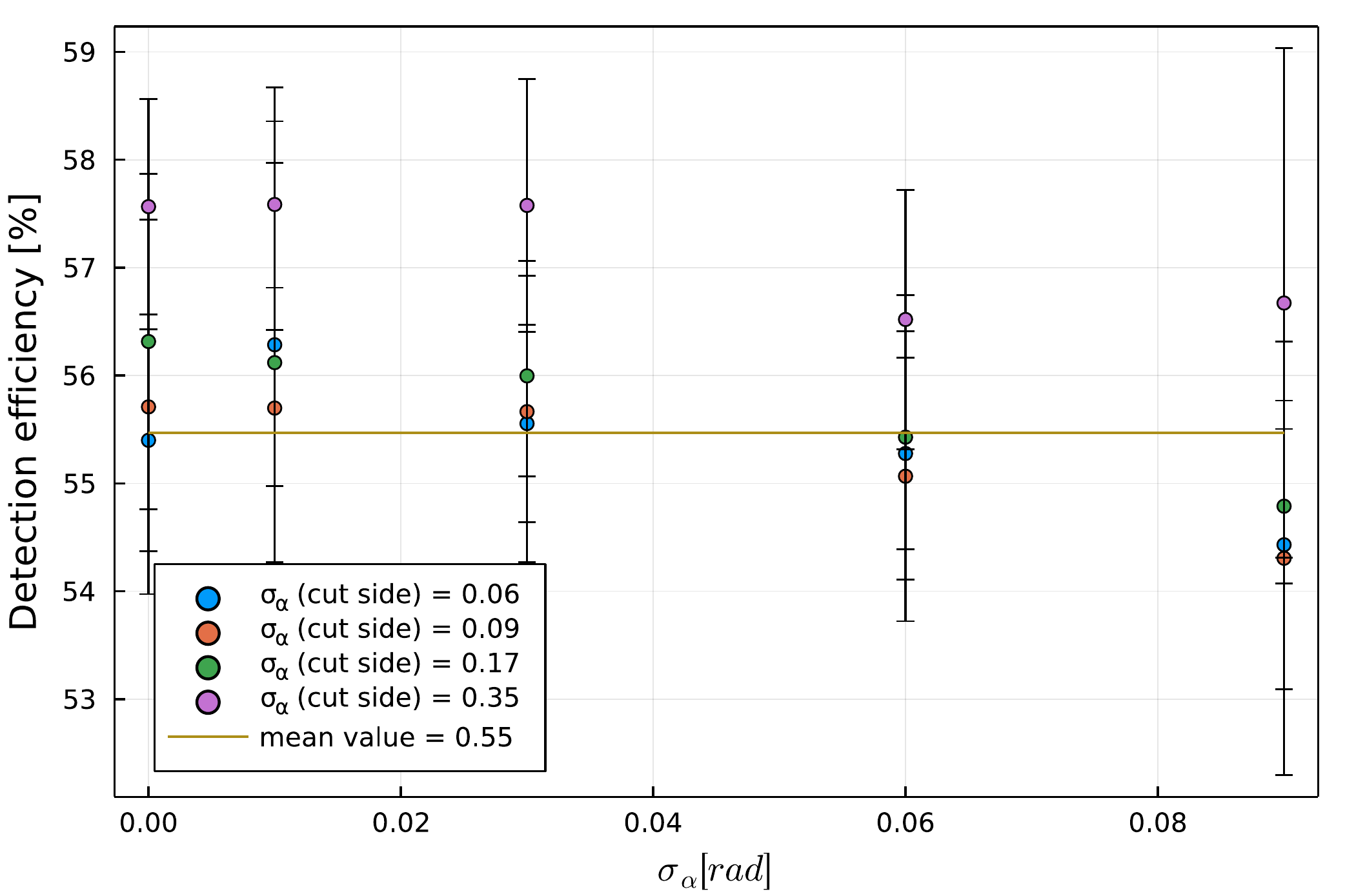}
    \caption{Left: Effect of the simulated light yield of PEN on the detection efficiency. A quadratic relation is used to fit the detection efficiency as a function of the simulated light yield. Right: Effect of the PEN surface roughness on the detection efficiency for a given  light yield. For each value of surface roughness of the molded sides ($x$-axis) four values of surface roughness were simulated for the lateral sides ($\sigma_\alpha$ of cut side), indicated with different colors in the plot. The mean value is computed taking the three lower values of surface roughness in the cut sides into account.   }
    \label{fig:pen_ly_fit}
\end{figure}

Finally, the response of the PEN samples was simulated using different values of light yield. These simulations were used to determine the detection efficiency as a function of light yield.
\begin{equation}
    \eta = \frac{\text{number of detected photons (data)}}{\text{ simulated number of detected photons (for a given light yield)}}
\end{equation}
with the "number of detected photons (data)" being fixed by the data. This implies that for a higher value of light yield used in the simulations a smaller value of $\eta$ is expected.  Figure~\ref{fig:pen_ly_fit} (Left) shows this relation for different assumptions of light yield. 
A light yield of 5437~photons/MeVe$^-$ best matches the calibrated detection efficiency of $\eta = 62$\% previously determined with the EJ-200 calibration samples.
Figure~\ref{fig:pen_ly_fit}~(Right) shows the effect of the surface roughness  on the determination of the detection efficiency for a given value of light yield. 

Taking into account the uncertainties on the different parameters used as input for this study, the total systematic uncertainty was estimated as
\begin{equation}
\sigma^2_{\text{PEN LY}}=\sigma^2_{\text{r-EJ200}}+\sigma^2_{\text{LY-EJ200}}+\sigma^2_{\text{r-PEN}}+\sigma^2_{\text{b-PEN}}+\sigma^2_{\text{setup}}
\end{equation}
where $\sigma_{\text{r-EJ200}}=$2\% and $\sigma_{\text{r-PEN}}=$1\% take into account the effects of the roughness of the EJ200 and PEN samples respectively, $\sigma_{\text{LY-EJ200}}=$2\% accounts for the uncertainty of the detection efficiency calibration coming from the uncertainty of the light yield of the EJ200 scintillator, $\sigma_{\text{b-PEN}}=$1\% stands for the uncertainty of the bulk absorption length, and $\sigma_{\text{setup}}=$1\% accounts for effects of alignment reproducibility of the measurements. A  total systematic uncertainty of $\sigma_{\text{PEN LY}}=$ 4\% ($\sim$220~photons/MeVe$^{-}$) was obtained.

\section{Light output and self-vetoing efficiency of the \textsc{Legend}-$200$ PEN holders}
\label{S:efficiency}

The light output of the \textsc{Legend}-$200$ PEN holders in the final operational conditions of the \textsc{Legend}-$200$ setup will depend on the shape and size of each holder. In addition, the different components mounted on the holders as well as the configuration of the light readout system will also impact the detection and self-vetoing efficiency.
While this light output can be simulated using the optical parameters reported in this work, having measurements under well controlled conditions are also crucial to compare and understand the light propagation and collection.

\subsection{The setup}
\begin{figure}[h]
    \centering
    \includegraphics[width=0.7\textwidth]{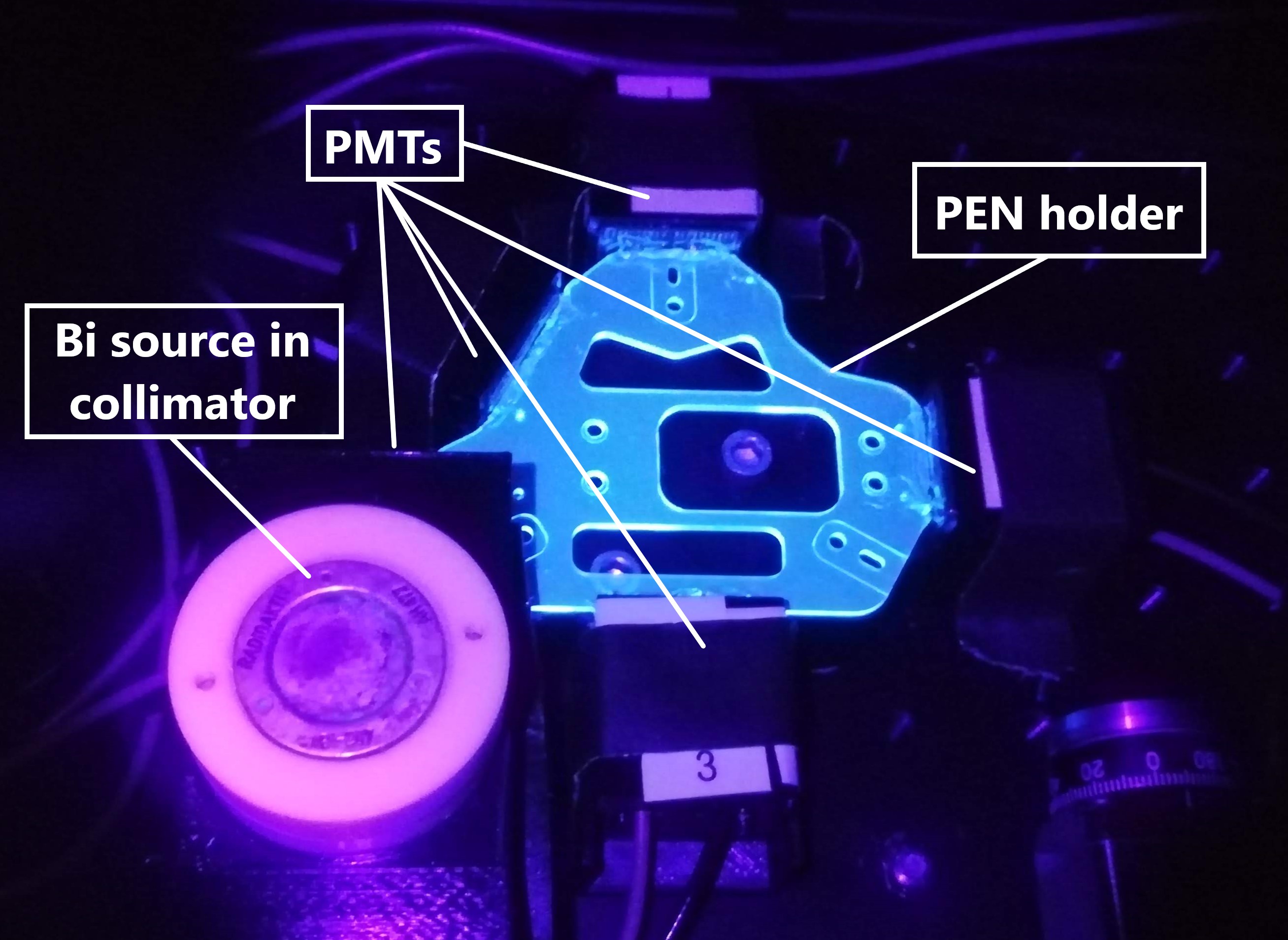}
    \caption{\textsc{Legend}-$200$ holder placed in a 3D support structure designed to fit its shape. The PMTs are coupled to the holder using optical grease.}
    \label{fig:output_l200_holders}
\end{figure}

In order to determine the light output and self-vetoing efficiency of the final PEN \textsc{Legend}-200 holders, PMTs were coupled to the flat lateral surfaces of the holders as can be seen in Figure~\ref{fig:output_l200_holders}. 3D printed support structures specially designed to fit the shape of each type of holder were used. These support structures allowed the placement of the PMTs to within 0.1~mm and therefore enable reproducible measurements. 
The optical grease, EJ-550, between the holders and the PMTs ensured a good optical coupling. Finally, the holders were excited using collimated electrons from the $^{207}$Bi source using the trigger system described previously.
The PEN holders were scanned along both axes in steps of 1~mm. In each position, 30 seconds of data were recorded. For a complete scan of each holder about 3 days of data taking were needed.

At the beginning of each scan, the PMTs were calibrated by their SPE response using the protocol described in subsection~\ref{subsec:pmt-calibration}. The goal of this calibration was to verify and decrease uncertainties caused by possible instabilities of the SPE response. However, it was found that the gain is very stable among the different measurements, and only sub-percent variations were observed.

\subsection{Light output and self-vetoing efficiency}

\begin{figure}[h]
    \centering
    \includegraphics[width=0.45\textwidth,trim={1cm 0 0cm 0}]{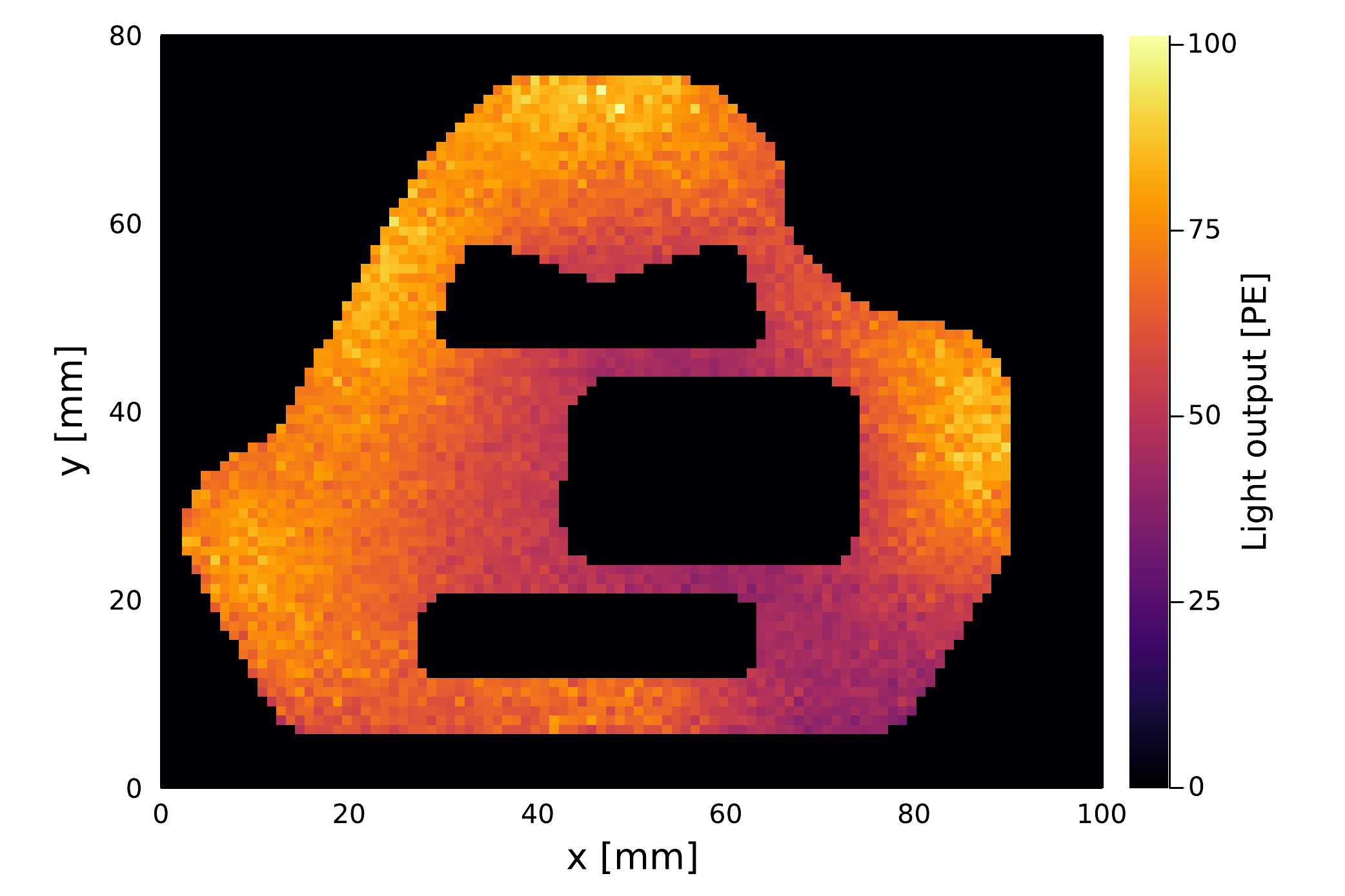}
    \includegraphics[width=0.45\textwidth,trim={1cm 0 0cm 0}]{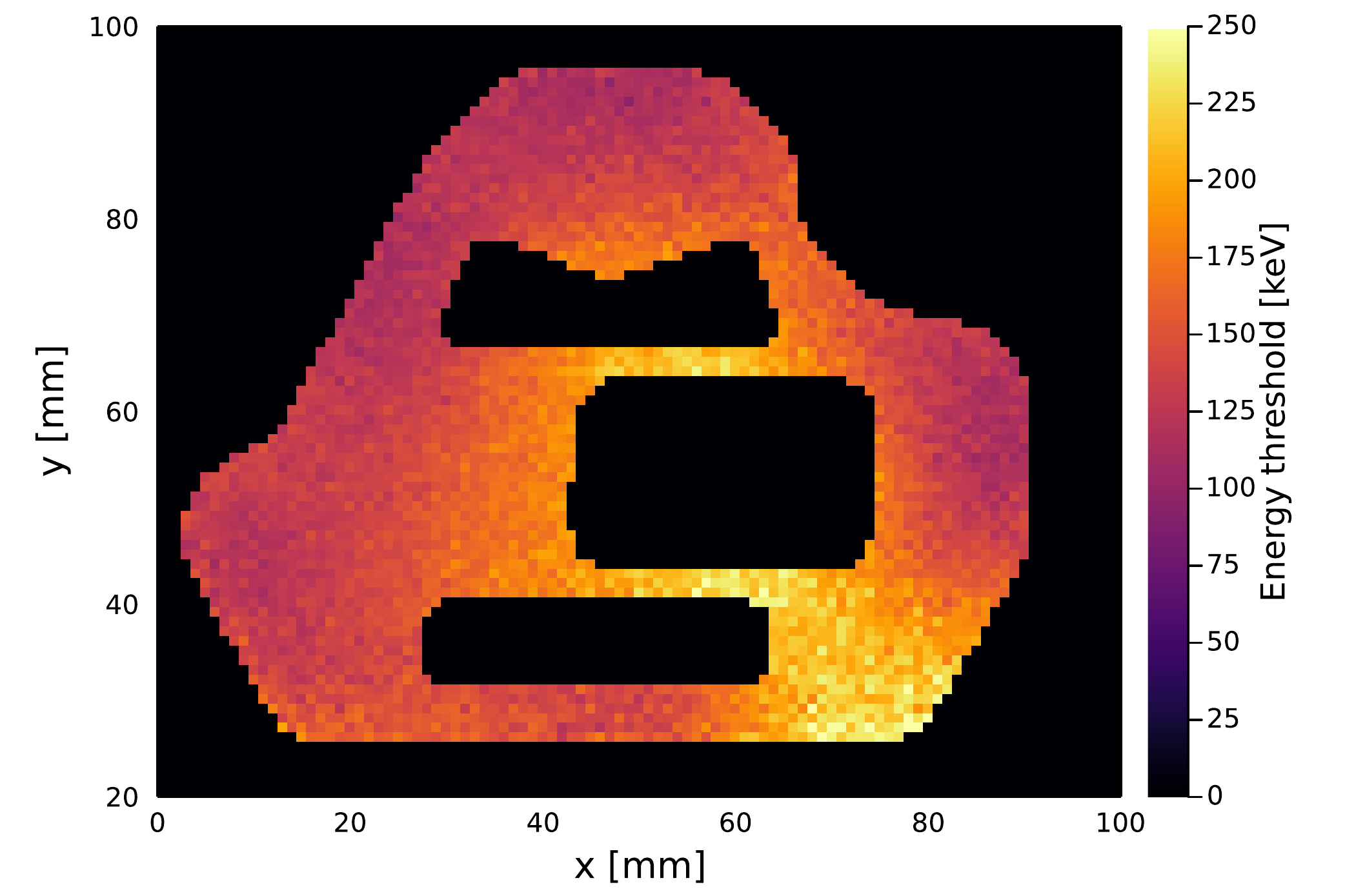}
    \caption{Left: Mean detected photons per event using electrons from the $^{207}$Bi source. For positions close to the PMTs around 100 p.e. are detected on average.
    Right: Required deposited energy to achieve a 5 $\sigma$ detection efficiency of at least 2 p.e. using a \textsc{Legend}-$200$ holder in this setup (see Figure~\ref{fig:output_l200_holders}). The lower threshold correspond to the regions close to the PMTs positions. The positions of the PMTs can be seen in Figure~\ref{fig:output_l200_holders}. }
    \label{fig:l200_holder_efficiency}
\end{figure}

To determine the light output, the charge of each PMT was converted to number of detected photons and then the total number of detected photons was found by summing the output of 
 all the PMTs coupled to the holder.
Figure~\ref{fig:l200_holder_efficiency}~(Left) shows the mean number of detected photons as a function of source position. 
Since the thickness of one holder is not enough to completely absorb the electrons from the $^{207}$Bi source, data with two stacked holders were taken for calibration purposes.
A calibration factor to convert the number of detected photons to deposited energy was found using the $\sim$1~MeV peak that could be identified using data with two stacked holders. 
Using this calibration factor the number of detected photons can be translated into energy deposited.

The self-vetoing efficiency of the \textsc{Legend}-$200$ holders depends on the optical properties of the material, the amount of deposited energy, type of particle and  the setup.
This study was performed assuming that the detected $\sim$1~MeV peak using the $^{207}$Bi source is Poisson distributed around its position. Therefore, by integrating the Poisson probability distribution from 0 detected photons to a certain amount, the probability of detecting at least that amount of photons for an energy deposition can be calculated. Assuming a linear energy dependency, this can be estimated for arbitrary energy depositions. Furthermore, by fixing the amount of photons that have to be detected, the minimum needed energy deposition to detect that amount of photons can be calculated.
In this way the minimum deposited energy required to have a 5 $\sigma$ detection probability of at least two photons in this setup was determined. Figure~\ref{fig:l200_holder_efficiency}~(Right) shows this estimated energy  as a function of $^{207}$Bi source position above the holder.

In general, it was found that for positions close to the PMTs, energy deposits of above 50 keV can be detected (>5$\sigma$). For some areas on the holder a minimum energy of $\sim$250~keV is required for a 5$\sigma$ detection.
The strong correlation with the position of the PMTs is explained because of geometrical effects but also it is a consequence of the bulk absorption length of PEN of the order of 60~mm. Hence, for events located far from the PMTs a significant reduction in the light output is produced. This effect is nevertheless tolerable for small components like the \textsc{Legend}-$200$ holders. 

\section{Conclusions \& Outlook}
\label{S:conlusion}
PEN is an innovative scintillating polymer material for use as an active structural material in low-background physics experiments. After a successful production of low background PEN tiles, support holders for deployment in the \legend-$200$ experiment were produced. These PEN holders will be used to mount the Ge detectors that are operated in liquid argon.
In this study, the surface of the \legend-$200$ holders has been characterized and it was found that high quality surfaces have been obtained. 
Thus, the molded surfaces of the PEN holders can be considered as having the surface quality of polished materials.
In the \legend-$200$ setup, this has the advantage that light from liquid argon produced in between the Ge detectors will be efficiently collected and guided (due to the difference in refractive index between PEN and LAr) to the lateral sides where the WLS optical fibers are placed. In this way, the use of  PEN holders will optimize light collection and the discrimination of backgrounds originating in the surrounding of the Ge detectors.

The main optical properties of the \legend-$200$ PEN production batch have been measured and are summarized in Table \ref{tab:parameters}. Detailed Geant4 simulations of the setups used for this optical characterization have been developed. Using the reported values of the different optical parameters, an excellent agreement between data and Monte Carlo simulations has been achieved. 

\begin{table}[h]
    \centering
    \begin{tabular}{ |c|c| } 
 \hline
 Light yield & $(5440\pm 220)$ photons/MeVe$^{-}$  \\ 
 \hline
 Time constant & $(25.3 \pm 0.2)$ ns \\ 
 \hline
 Bulk absorption length & $(62 \pm 3)$ mm at 450 nm  \\ 
 \hline
 $\sigma_\alpha$ of molded surfaces & < 0.1 rad  \\
 \hline 
 Emission spectrum maximum & $(440 \pm 3)$ nm  \\ 
 \hline
\end{tabular}
    \caption{Main optical properties of \legend-$200$ PEN scintillator batch.}
    \label{tab:parameters}
\end{table}

The final \legend-$200$ holders have been optically characterized and an estimation of the minimum energy needed to be deposited for their self-vetoing capabilities within the PMT setup used in this study has been performed. It was found that a self-vetoing efficiency (>5$\sigma$ detection) for events depositing a minimum energy between 50 and of 250 keV (depending on the position) is reached.
Even though these results are setup dependent, these open the window - using detailed MC simulations - for a precise estimation of the background discrimination and self-vetoing efficiency of these PEN holders in the final operational conditions of the \legend-$200$ setup. 
Moreover, these results can be used to study the benefits of using PEN in similar low-background physics experiments.
Finally, further R\&D is being carried out for a potential wider application of PEN in the \legend-$1000$ experiment.

\acknowledgments
The PEN holders used for this study were produced as part of the \legend R\&D program within the PEN consortium. 
We gratefully acknowledge all the members of PEN consortium and the \legend collaboration for their support. 
Felix Fischer, Andreas Leonhardt, and Luis Manzanillas  are supported by the Deutsche Forschungsgemeinschaft (SFB1258).

\bibliographystyle{JHEP}
\bibliography{bib}

\providecommand{\href}[2]{#2}\begingroup\raggedright\begin{thebibliography}{10}

\bibitem{Heindl:2010zz}
T.~Heindl, T.~Dandl, M.~Hofmann, R.~Krucken, L.~Oberauer, W.~Potzel et~al.,
  \emph{{The scintillation of liquid argon}},
  \href{http://dx.doi.org/10.1209/0295-5075/91/62002}{\emph{EPL} {\bfseries 91}
  (2010) 62002}, [\href{https://arxiv.org/abs/1511.07718}{{\ttfamily
  1511.07718}}].

\bibitem{Agostini:2017hit}
{\scshape GERDA} collaboration, M.~Agostini et~al., \emph{{Upgrade for Phase II
  of the Gerda experiment}},
  \href{http://dx.doi.org/10.1140/epjc/s10052-018-5812-2}{\emph{Eur. Phys. J.
  C} {\bfseries 78} (2018) 388},
  [\href{https://arxiv.org/abs/1711.01452}{{\ttfamily 1711.01452}}].

\bibitem{LEGEND:2021bnm}
{\scshape LEGEND} collaboration, N.~Abgrall et~al., \emph{{The Large Enriched
  Germanium Experiment for Neutrinoless $\beta\beta$ Decay}: {LEGEND-1000
  Preconceptual Design Report}},
  \href{https://arxiv.org/abs/2107.11462}{{\ttfamily 2107.11462}}.

\bibitem{Efremenko:2019xbs}
Y.~Efremenko et~al., \emph{{Use of poly(ethylene naphthalate) as a self-vetoing
  structural material}},
  \href{http://dx.doi.org/10.1088/1748-0221/14/07/P07006}{\emph{JINST}
  {\bfseries 14} (2019) P07006},
  [\href{https://arxiv.org/abs/1901.03579}{{\ttfamily 1901.03579}}].

\bibitem{Efremenko:2021olf}
Y.~Efremenko et~al., \emph{{Production and validation of scintillating
  structural components from low-background Poly(ethylene naphthalate)}},
  \href{http://dx.doi.org/10.1088/1748-0221/17/01/P01010}{\emph{JINST}
  {\bfseries 17} (2022) P01010},
  [\href{https://arxiv.org/abs/2110.12791}{{\ttfamily 2110.12791}}].

\bibitem{Kuzniak:2020oka}
M.~Ku\'zniak and A.~M. Szelc, \emph{{Wavelength Shifters for Applications in
  Liquid Argon Detectors}},
  \href{http://dx.doi.org/10.3390/instruments5010004}{\emph{Instruments}
  {\bfseries 5} (2020) 4}, [\href{https://arxiv.org/abs/2012.15626}{{\ttfamily
  2012.15626}}].

\bibitem{Abraham:2021otn}
Y.~Abraham et~al., \emph{{Wavelength-Shifting Performance of Polyethylene
  Naphthalate Films in a Liquid Argon Environment}},
  \href{https://arxiv.org/abs/2103.03232}{{\ttfamily 2103.03232}}.

\bibitem{Kuzniak:2018dcf}
M.~Ku\'zniak, B.~Broerman, T.~Pollmann and G.~R. Araujo, \emph{{Polyethylene
  naphthalate film as a wavelength shifter in liquid argon detectors}},
  \href{http://dx.doi.org/10.1140/epjc/s10052-019-6810-8}{\emph{Eur. Phys. J.
  C} {\bfseries 79} (2019) 291},
  [\href{https://arxiv.org/abs/1806.04020}{{\ttfamily 1806.04020}}].

\bibitem{Araujo:2021buv}
G.~R. Araujo et~al., \emph{{R\&D of Wavelength-Shifting Reflectors and
  Characterization of the Quantum Efficiency of Tetraphenyl Butadiene and
  Polyethylene Naphthalate in Liquid Argon}},
  \href{https://arxiv.org/abs/2112.06675}{{\ttfamily 2112.06675}}.

\bibitem{Boulay:2021njr}
M.~G. Boulay et~al., \emph{{Direct comparison of PEN and TPB wavelength
  shifters in a liquid argon detector}},
  \href{http://dx.doi.org/10.1140/epjc/s10052-021-09870-7}{\emph{Eur. Phys. J.
  C} {\bfseries 81} (2021) 1099},
  [\href{https://arxiv.org/abs/2106.15506}{{\ttfamily 2106.15506}}].

\bibitem{Abramov:2019hhh}
{\scshape GERDA} collaboration, M.~Agostini et~al., \emph{{Modeling of GERDA
  Phase II data}}, \href{http://dx.doi.org/10.1007/JHEP03(2020)139}{\emph{JHEP}
  {\bfseries 03} (2020) 139},
  [\href{https://arxiv.org/abs/1909.02522}{{\ttfamily 1909.02522}}].

\bibitem{Abt:2020pwk}
I.~Abt et~al., \emph{{Usage of PEN as self-vetoing structural material in low
  background experiments}},
  \href{http://dx.doi.org/10.22323/1.390.0163}{\emph{PoS} {\bfseries ICHEP2020}
  (2021) 163}, [\href{https://arxiv.org/abs/2011.08983}{{\ttfamily
  2011.08983}}].

\bibitem{Manzanillas:2022pat}
L.~Manzanillas et~al., \emph{{Usage of PEN as self-vetoing structural material
  in the LEGEND experiment}},
  \href{http://dx.doi.org/10.1088/1748-0221/17/03/C03031}{\emph{JINST}
  {\bfseries 17} (2022) C03031}.

\bibitem{hamamatsuPMT}
``Manual of the h11934 pmt series.''
  \url{https://www.hamamatsu.com/content/dam/hamamatsu-photonics/sites/documents/99_SALES_LIBRARY/etd/R11265U_H11934_TPMH1336E.pdf}.

\bibitem{PhysRev.99.695}
D.~E. Alburger and A.~W. Sunyar, \emph{Decay of ${\mathrm{bi}}^{207}$},
  \href{http://dx.doi.org/10.1103/PhysRev.99.695}{\emph{Phys. Rev.} {\bfseries
  99} (Aug, 1955) 695--702}.

\bibitem{tableisotopes}
M.~Be and V.~Chiste, \emph{Table de Radionucl\'eide, Bi-207}.
\newblock Paris France, 2009.

\bibitem{Abreu:2018ajc}
{\scshape SoLid} collaboration, Y.~Abreu et~al., \emph{{Optimisation of the
  scintillation light collection and uniformity for the SoLid experiment}},
  \href{http://dx.doi.org/10.1088/1748-0221/13/09/P09005}{\emph{JINST}
  {\bfseries 13} (2018) P09005},
  [\href{https://arxiv.org/abs/1806.02461}{{\ttfamily 1806.02461}}].

\bibitem{eljen}
``{ELJEN TECHNOLOGY:} general purpose plastic scintillators.''
  \url{https://eljentechnology.com/products/plastic-scintillators/ej-200-ej-204-ej-208-ej-212}.

\bibitem{bezanson2017julia}
J.~Bezanson, A.~Edelman, S.~Karpinski and V.~B. Shah, \emph{Julia: A fresh
  approach to numerical computing}, {\emph{SIAM review} {\bfseries 59} (2017)
  65--98}.

\bibitem{Bauer:2011ne}
C.~Bauer et~al., \emph{{Qualification Tests of 474 Photomultiplier Tubes for
  the Inner Detector of the Double Chooz Experiment}},
  \href{http://dx.doi.org/10.1088/1748-0221/6/06/P06008}{\emph{JINST}
  {\bfseries 6} (2011) P06008},
  [\href{https://arxiv.org/abs/1104.0758}{{\ttfamily 1104.0758}}].

\bibitem{Anthony_2018}
M.~Anthony, E.~Aprile, L.~Grandi, Q.~Lin and R.~Saldanha,
  \emph{Characterization of photomultiplier tubes with a realistic model
  through {GPU}-boosted simulation},
  \href{http://dx.doi.org/10.1088/1748-0221/13/02/t02011}{\emph{Journal of
  Instrumentation} {\bfseries 13} (feb, 2018) T02011--T02011}.

\bibitem{Agostinelli:2002hh}
{\scshape GEANT4} collaboration, S.~Agostinelli et~al., \emph{{GEANT4--a
  simulation toolkit}},
  \href{http://dx.doi.org/10.1016/S0168-9002(03)01368-8}{\emph{Nucl. Instrum.
  Meth. A} {\bfseries 506} (2003) 250--303}.

\bibitem{unified_model_citation}
C.~Moisan, E.~M. Hoskinson, A.~Levin and D.~Vozza, \emph{{Public domain
  platform to model scintillation counters for gamma-ray imaging
  applications}},  in \emph{Hard X-Ray and Gamma-Ray Detector Physics, Optics,
  and Applications} (R.~B. Hoover and F.~P. Doty, eds.), vol.~3115, pp.~21 --
  29, International Society for Optics and Photonics, SPIE, 1997.
\newblock \href{http://dx.doi.org/10.1117/12.277694}{DOI}.

\bibitem{SurfacesGeant4}
A.~Levin and C.~Moisan, \emph{A more physical approach to model the surface
  treatment of scintillation counters and its implementation into detect},  in
  \emph{1996 IEEE Nuclear Science Symposium. Conference Record}, vol.~2,
  pp.~702--706 vol.2, 1996.
\newblock \href{http://dx.doi.org/10.1109/NSSMIC.1996.591410}{DOI}.

\bibitem{pennakamura}
H.~Nakamura, Y.~Shirakawa, S.~Takahashi and H.~Shimizu, \emph{Evidence of
  deep-blue photon emission at high efficiency by common plastic}, {\emph{EPL}
  {\bfseries 95} (2011) }.

\bibitem{Bilki:2019lep}
J.~Wetzel, N.~Bostan, O.~K. K\"oseyan, E.~Tiras and B.~Bilki,
  \emph{{Scintillation timing characteristics of common plastics for radiation
  detection excited with 120 GeV protons}},
  \href{http://dx.doi.org/10.3906/fiz-1912-9}{\emph{Turk. J. Phys.} {\bfseries
  44} (2020) 437--441}, [\href{https://arxiv.org/abs/1912.11342}{{\ttfamily
  1912.11342}}].

\bibitem{Nakamura_2019}
H.~Nakamura and K.~Mori, \emph{Time response of poly (ethylene naphthalate)
  light emission to charged particles},
  \href{http://dx.doi.org/10.1088/1402-4896/ab247b}{\emph{Physica Scripta}
  {\bfseries 94} (aug, 2019) 105302}.

\bibitem{Marchi:2019moe}
T.~Marchi et~al., \emph{{Optical properties and pulse shape discrimination in
  siloxane-based scintillation detectors}},
  \href{http://dx.doi.org/10.1038/s41598-019-45307-8}{\emph{Sci. Rep.}
  {\bfseries 9} (2019) 9154}.

\bibitem{Hackett:2022xnk}
B.~Hackett et~al., \emph{{Light Response of Poly(ethylene 2,6-napthalate) to
  Neutrons}},  \href{https://arxiv.org/abs/2204.02866}{{\ttfamily 2204.02866}}.

\end{thebibliography}\endgroup
\end{document}